\documentclass[12pt]{article}
\usepackage{amsmath,amsfonts,graphicx,epsfig,color}
\def\no{{}^\circ_{\circ}}
\begin{document}

\baselineskip=.22in
\renewcommand{\baselinestretch}{1.2}
\renewcommand{\theequation}{\thesection.\arabic{equation}}
\makeatletter\@addtoreset{equation}{section}
\begin{flushright}
TCDMATH 08--02
\end{flushright}

\vspace{3mm}

\begin{center}
{{{\Large \bf Negative-Tension Branes and Tensionless $\frac{1}{2}$Brane\\
in\\ Boundary Conformal Field Theory}}\\[12mm]
{Akira Ishida${}^{(1)}$, Chanju Kim${}^{(2)}$,
Yoonbai Kim${}^{(1)}$, O-Kab Kwon${}^{(3)}$}\\[4mm]
{\it ${}^{(1)}$Department of Physics and BK21 Physics Research Division,
\\ Sungkyunkwan University, Suwon 440-746, Korea}\\
{\tt ishida@skku.edu,~~yoonbai@skku.edu}\\[5mm]
{\it ${}^{(2)}$Department of Physics, Ewha Womans University,
Seoul 120-750, Korea}\\
{\tt cjkim@ewha.ac.kr}\\[5mm]
{\it ${}^{(3)}$School of Mathematics, Trinity College Dublin,
Ireland}}\\
{\tt okabkwon@maths.tcd.ie}

\end{center}
\vspace{10mm}

\begin{abstract}
In the framework of boundary conformal field theory we consider a flat
unstable D$p$-brane in the presence of a large constant electromagnetic field.
Specifically, we study the case that the electromagnetic field
satisfy the following three conditions:
(i) a constant electric field is turned on along the $x^1$ direction
($E_{1}\ne 0$); (ii) the determinant of the matrix $(\eta + F)$ is
negative so that it lies in the physical region ($-\det (\eta + F)>0$);
(iii) the 11-component of its cofactor is positive
to the large electromagnetic field.
In this case, we identify exactly
marginal deformations depending on the spatial coordinate $x^1$. They
correspond to tachyon profiles of hyperbolic sine, exponential, and
hyperbolic cosine types. Boundary states are constructed for these
deformations by utilizing T-duality approach and also by directly solving
the overlap conditions in BCFT. The exponential type deformation gives
a tensionless half brane connecting the perturbative string vacuum and one
of the true tachyon vacua, while the others have negative tensions.
This is in agreement with the results obtained in other approaches.
\end{abstract}


\newpage

\section{Introduction}
\label{section1}

Type IIA (IIB) string theory supports even-dimensional (odd-dimensional)
stable BPS D-branes and odd-dimensional (even-dimensional) unstable
nonBPS D-branes~\cite{Sen:2004nf}.
The physics of unstable D-branes involves various
nonperturbative aspects of string theory.
Specifically, two representative examples are the tachyon solitons interpreted
as lower-dimensional D-branes~\cite{Sen:1998tt,Sen:1998ex}
and the rolling tachyon describing homogeneous real-time decay process
of the D-brane~\cite{Sen:2002nu,Sen:2002in}.

The instability of a nonBPS D$p$-brane in superstring theory
or a D$p$-brane in bosonic string theory
results in the appearance of a tachyonic degree.
In the context of boundary conformal field theory (BCFT),
the tachyon vertex operator with a single spatial dependence
which represents the exactly marginal deformation
is given by a sinusoidal function with a single
multiplicative parameter.
When the parameter has the value 1/2, the deformation is interpreted as
an array of D$(p-1)$-branes in bosonic string
theory or as a periodic array of a pair of a D$(p-1)$-brane and
a $\bar{\rm D}(p-1)$-brane
in superstring theory~\cite{Sen:1998tt,Sen:1998ex,Sen:1999mh}.
The homogeneous rolling tachyon in BCFT is described
by introducing a marginal deformation corresponding to the tachyon
profiles of hyperbolic sine, exponential, or hyperbolic cosine type.
The physical quantities like the energy-momentum tensor suggest
real-time decay of an unstable D-brane
when the tachyon is displaced from the maximum of the tachyon potential
and rolls down towards its minimum.

Both homogeneous rolling tachyons and lower-dimensional D-branes
from an unstable D-brane have also been studied in the context of
effective field theories (EFTs) such as Dirac-Born-Infeld (DBI) type
EFT \cite{Sen:2002an,Lambert:2003zr,Kim:2003in,Brax:2003rs}
and boundary string field theory (BSFT) EFT
\cite{Gerasimov:2000zp,Sugimoto:2002fp,Kim:2006mg}.
Compared with the BCFT approach, physical quantities such as
energy-momentum tensor obtained from these EFTs are qualitatively
the same as, but slightly different from that in BCFT
\cite{Sen:2002an}--\cite{Kim:2006mg}.
For the case of the half S-brane with the exponential type of the tachyon
profile~\cite{Maloney:2003ck} which is a special
case of homogeneous rolling tachyons,
the energy-momentum tensor based on the formula in Ref.~\cite{Sen:2002in}
coincides exactly with that of DBI type effective
action with 1/cosh potential~\cite{Kutasov:2003er}.

When the fundamental strings exist in the worldvolume of unstable
D$p$-brane, they couple to the second-rank antisymmetric tensor field
(or equivalently to the electromagnetic field strength tensor on the
D-brane~\cite{Witten:1995im})
and the string current density
is given by the Lorentz-covariant conjugate momentum of U(1) gauge
field~\cite{Yi:1999hd,Gibbons:2000hf,Sen:2000kd}. Then one may study
the effect of the electromagnetic field. For rolling tachyons,
the three types of deformations mentioned above are not changed
by the presence of
constant electric~\cite{Mukhopadhyay:2002en,Gibbons:2002tv,Ishida:2002fr},
or both electric and magnetic~\cite{Rey:2003xs,Kim:2003he} fields
as far as they satisfy the physical condition, $-\det(\eta+F)>0$ where
$F$ denotes the electromagnetic field strength tensor.

The situation is more interesting for the case of tachyon kinks which
are identified as lower-dimensional D-branes of codimension one.
The spectrum of the tachyon kink is not changed
when the constant electric field is turned on with keeping $-\det(\eta+F)>0$;
only the period of D$(p-1){\bar {\rm D}} (p-1)$ in the array becomes large
as the electric field increases. However, when the electric field eventually
reaches the critical value for which the determinant vanishes, the period
becomes infinite and we obtain a single regular BPS tachyon kink with
constant electric flux. It is interpreted
as a thick BPS D$(p-1)$-brane in the background of fundamental string charge
density. This has been checked in various languages including DBI
EFT~\cite{Kim:2003in,Kim:2003ma}, BCFT~\cite{Sen:2003bc},
noncommutative field theory (NCFT)~\cite{Banerjee:2004cw}, and
BSFT~\cite{Kim:2006mg}.

In the presence of both constant electric and constant magnetic fields
in an unstable D$p$-brane with $p\ge 2$, it turns out that
other types of deformations are possible. This is because
the 11-component of the cofactor $C^{11}$ of
$(\eta + F)_{\mu\nu}$ can have either negative or positive value while
keeping $-\det (\eta + F) > 0$. (Here, $x^1$ denotes the coordinate on
which the tachyon depends.)
For small electromagnetic fields the cofactor $C^{11}$ is negative.
In this case the species of tachyon kinks are essentially the same
as those without electromagnetic field.
On the other hand, when $p\ge 2$, electromagnetic fields can take large
values for which $C^{11}$ becomes positive while
maintaining the condition $-\det (\eta + F) > 0$.
In this case, three new codimension-one objects are supported,
which correspond to tachyon profiles of hyperbolic
sine, exponential, and hyperbolic cosine types.
These objects have been obtained in aforementioned EFTs
including DBI EFT~\cite{Kim:2003in,Kim:2003ma}, NCFT~\cite{Kim:2004xn},
and BSFT~\cite{Kim:2006mg}. They, however, have not yet been reproduced in
the context of BCFT for type II superstring theory. The purpose of this paper
is to analyze these three kinks in the context of BCFT.

In section~\ref{section2}, we describe new tachyon vertices of
hyperbolic sine, exponential, and hyperbolic cosine types
with the dependence on a single
spatial coordinate, and show that they give marginal deformations in the
context of BCFT. In addition, these tachyon profiles are obtained
as static solutions of the linearized tachyon equation in the background of
a nontrivial open string metric and a noncommutativity parameter of
open string field theory (OSFT).
In section \ref{section3} we construct the boundary states corresponding to
the marginal deformations given in section~\ref{section2}.
We utilize Lorentz transformation and T-duality in subsection
\ref{subsection3-1} while the overlap condition is directly solved
in subsection \ref{subsection3-2} to construct the boundary states.
In section~\ref{section4} we read the corresponding physical quantities,
specifically the energy-momentum tensor
$T_{\mu\nu}$ and the fundamental string current density $\Pi_{\mu\nu}$.
For the case of hyperbolic sine and hyperbolic cosine types of tachyon
profiles, they are slightly different from those in EFTs as
expected~\cite{Kim:2003in,Kim:2003ma,Kim:2004xn},
but, for the exponential type, they coincide exactly with the results of EFTs.
They are interpreted as negative tension branes
for hyperbolic sine and cosine profiles and tensionless half brane
for exponential profile in the huge constant background of positive
energy density. We conclude in section~\ref{section5}.
In the appendix, we give an alternative derivation of the energy-momentum
tensor for the exponential type deformation by calculating the partition
function of the worldsheet action following Ref.~\cite{Larsen:2002wc}.

\section{New Tachyon Vertices as Exactly Marginal Deformations}
\label{section2}

In this section we show that there exist three new tachyon vertices
as marginal deformations in the presence of the constant electromagnetic field.
They are hyperbolic sine, hyperbolic cosine, and exponential types
which depend on a single spatial coordinate.
We shall show this first in the scheme of BCFT and then in the context
of linearized OSFT.

In the BCFT description of bosonic string theory,
the worldsheet action of a D$p$-brane in the
presence of a background U(1) gauge field $A_\mu$ is given by\footnote{
Throughout this paper we use the $\alpha'=1$ unit.}
\begin{equation}\label{SBCFT}
S_{\rm BCFT} = \frac1{2\pi}\int_{\Sigma}d^2w\,
\partial X^\mu\bar\partial X_\mu
- \frac{i}{2\pi} \int_{\partial \Sigma}dt\, A_\mu (X) \partial_t X^\mu,
\end{equation}
where $\Sigma$ denotes the worldsheet and $t$ parametrizes the boundary
of the worldsheet. It satisfies the boundary condition
\begin{equation} \label{bnd1}
\left( \partial_n X^\mu
 + iF^\mu{}_\nu \partial_t X^\nu \right) \mid_{\partial\Sigma} = 0,
\end{equation}
where $\partial_n$ and $\partial_t$ denote respectively the normal and
tangential derivatives at the boundary $\partial\Sigma$.
The dynamics of an unstable D$p$-brane is described by introducing
a conformally invariant boundary interaction to the worldsheet action,
\begin{equation}
S_T =\int_{\partial \Sigma} dt\, T(X). \label{ST}
\end{equation}

When there is no background gauge field ($A_\mu=0$), it is well-known
that for any spatial direction $X$ the operator
\begin{equation} \label{cosx}
T(X) = \lambda \cos X
\end{equation}
is exactly marginal and has been used to describe lower-dimensional
D-branes \cite{Sen:1998tt,Sen:1998ex}.
The relevant boundary state is given by
\cite{Callan:1994ub,Polchinski:1994my,Recknagel:1998ih}
\begin{equation} \label{bstate}
|\mathcal{B}\rangle_X = \sum_{j = 0,1/2,1,\ldots} \sum_{m=-j}^j D_{m,-m}^j(R)
              | j;m,m \rangle\rangle,
\end{equation}
where $R$ is the SU(2) rotation matrix
\begin{equation}
R = \begin{pmatrix}
        \cos\pi\lambda & i\sin\pi\lambda \\
        i\sin\pi\lambda & \cos\pi\lambda
    \end{pmatrix},
\end{equation}
$D_{m,-m}^j(R)$ is the spin $j$ representation matrix of $R$
in the $J_z$ eigenbasis, and $ | j;m,m \rangle\rangle $ is the
Virasoro-Ishibashi state \cite{Ishibashi:1988kg} built over the primary state
$ |j;m,m\rangle = |j,m\rangle \overline{|j,m\rangle}$.

The temporal version of (\ref{cosx})
\begin{equation} \label{rolltach}
T(X^0) = \lambda \cosh X^0
\end{equation}
describes the dynamics of the rolling tachyon \cite{Sen:2002nu,Sen:2002in}.
For example the relevant boundary state for a D$p$-brane with the deformation
(\ref{rolltach}) is given by
\begin{equation}
|\mathcal{B} \rangle = |\mathcal{B} \rangle_{X^0}
  \otimes_{\mu=1}^p | N \rangle_{X^\mu}
  \otimes_{i=p+1}^{25} | D \rangle_{X^i}
 \otimes |\text{ghost} \rangle,
\end{equation}
where
\begin{align}
|N \rangle_{X^\mu}
  &= \exp\left[ -\sum_{n=1}^\infty \frac1n \alpha_{-n}^\mu \bar\alpha_{-n}^\mu
          \right]|0\rangle, \nonumber \\
|D \rangle_{X^i}
   &= \exp\left[ \sum_{n=1}^\infty \frac1n \alpha_{-n}^i \bar\alpha_{-n}^i
          \right]|0\rangle, \nonumber \\
|\text{ghost} \rangle
   &= \exp\left[ -\sum_{n=1}^\infty (\bar b_{-n} c_{-n} + b_{-n} \bar c_{-n})
          \right](c_0 + \bar c_0)c_1 \bar c_1 |0\rangle,
\end{align}
and $|{\cal B}\rangle_{X^0}$ is the boundary state (\ref{bstate}) in
the Wick-rotated variable $X\to-iX^0$.

In the presence of a constant background electric field
the operator (\ref{rolltach}) is modified by a Lorentz factor
\cite{Mukhopadhyay:2002en},
\begin{equation} \label{rolltach2}
T(X^0) = \lambda \cosh(\sqrt{1-E^2}\, X^0),
\end{equation}
where $E$ is the magnitude of the electric field. Rolling tachyons have
further been generalized to the case that both electric and magnetic fields
are turned on~\cite{Rey:2003xs,Kim:2003he}.

In this section we would like to discuss the exactly marginal deformation
$S_T$ (\ref{ST}) in the background of a general constant
electromagnetic field for which we take the symmetric gauge,
\begin{equation}
A_\mu = -\frac12 F_{\mu\nu}X^\nu. \label{SGa}
\end{equation}
As usual \cite{Sen:2002nu}, we first consider the Wick-rotated theory
obtained by the replacement $X^0 \rightarrow iX^0$ and make the inverse
Wick-rotation back to the Minkowski time later.

Under the deformed boundary condition (\ref{bnd1}),
the correlation function on the upper-half plane is obtained
as~\cite{Abouelsaood:1986gd},
\begin{align}
\langle X^\mu(w) X^\nu(w')\rangle
=&-\delta^{\mu\nu}\ln \mid w-w'\mid +\delta^{\mu\nu}\ln \mid w-\bar w'\mid
\nonumber\\
&-\bar G^{\mu\nu}\ln \mid w-\bar w'\mid^2
- \bar\theta^{\mu\nu}\ln \left(\frac{w-\bar w'}{\bar w - w'}\right).
\label{XX6}
\end{align}
Here $\bar G_{\mu\nu}$ and $\bar\theta^{\mu\nu}$ are the Wick-rotated
version of the open string metric $G^{\mu\nu}$ and
the noncommutativity parameter $\theta^{\mu\nu}$ given as
\begin{equation}\label{op}
G^{\mu\nu}=\left(\frac{1}{\eta+F}\right)^{\mu\nu}_{{\rm S}}
=\frac{C^{\nu\mu}_{{\rm S}}}{{\cal Y}_{p}}, \qquad
\theta^{\mu\nu}=\left(\frac{1}{\eta+F}\right)^{\mu\nu}_{{\rm A}}
=\frac{C^{\nu\mu}_{{\rm A}}}{{\cal Y}_{p}},
\end{equation}
where $C_{{\rm S}}^{\mu\nu}$ and $C_{{\rm A}}^{\mu\nu}$ are respectively
the symmetric and antisymmetric parts of $C^{\mu\nu}$,
the cofactor matrix of $(\eta+F)_{\mu\nu}$, and
${\cal Y}_{p}\equiv \det (\eta + F)$.
When $w'$ is on the boundary, (\ref{XX6}) reduces to
\begin{equation} \label{XX9}
X^\mu(w)X^\nu(t) \sim -\bar G^{\mu\nu} \ln |w-t|^2 -\bar\theta^{\mu\nu}
\ln \left(\frac{w-t}{\bar w -t}\right),
\end{equation}
where $t$ represents the boundary.
Since the OPE of the energy-momentum tensor
$T=-\partial X^\mu\partial X_\mu$ with the tachyon boundary
vertex operator $e^{ik\cdot X(t)}$ is
\begin{equation}
T(w) e^{ik\cdot X(t)} \sim \frac{\bar G^{\mu\nu}k_\mu k_\nu}
{(w-t)^2}\, e^{ik\cdot X(t)} + \frac{1}{w-t}\,\partial_t (e^{ik\cdot X(t)}),
\end{equation}
this operator becomes marginal when
\begin{equation}\label{const1}
\bar G^{\mu\nu}k_\mu k_\nu=1.
\end{equation}
In this paper we are interested in the operators which
depend on a single spatial coordinate.
For definiteness we take $k_\mu \propto \delta_{\mu1}$ and denote
$X=X^1$ for the sake of simplicity.
In this case (\ref{const1}) reduces to $ k_1^2 = 1/{\bar G^{11}}$.
Then the boundary operator is not only marginal but actually
exactly marginal \cite{Recknagel:1998ih} since the second term containing
the noncommutative parameter in (\ref{XX9}) plays no role.
After the inverse Wick rotation, the marginality condition then leads to
\begin{equation} \label{k1}
k_1^2 = \frac1{G^{11}}
  = \frac{{\cal Y}_{p}}{C^{11}}.
\end{equation}
As an example, let us consider an unstable D2-brane ($p=2$)
with a constant electromagnetic field
$F_{0i}=E_{i}$ (${\bf E}^{2}=E_{i}^{2}$) and $F_{12}=B$.
In this case, ${\cal Y}_{2}= -(1-{\bf E}^{2}+B^{2})$ and $C^{11} = -1+E_2^2$
so that the marginality condition (\ref{k1}) becomes
\begin{equation}\label{k12}
k_1^2=\frac{-(1 - {\bf E}^2 + B^2)}{-(1 - E_2^2)}.
\end{equation}

Before discussing the physical meaning of these marginal deformations,
we extend our analysis to the
superstring case with
worldsheet fermions, $\psi^{\mu}$ and $\bar\psi^{\mu}$,
$\mu=0,1,\ldots,9$.
The worldsheet action with a constant electromagnetic field is
\begin{equation}\label{Sw}
S_{{\rm w}}=\frac{1}{2\pi} \int_{\Sigma} d^2 z \left(
\partial X^{\mu}{\bar \partial}X_{\mu}
+\frac12\psi^{\mu}{\bar \partial}\psi_{\mu}
+\frac12\bar\psi^{\mu}\partial\bar\psi_{\mu}\right)
-\frac{i}{2\pi}\int_{\partial\Sigma} dt
\left(A_{\mu}\partial_tX^{\mu}- F_{\mu\nu}\Psi^{\mu}
\Psi^{\nu} \right),
\end{equation}
where the fermions in the boundary interaction always appear as
the following combination,
\begin{equation}
\Psi^\mu=\frac{1}{2}(\psi^\mu+\bar\psi^\mu).
\end{equation}
In addition to the boundary condition for bosonic degrees
(\ref{bnd1}), we impose that for fermionic degrees,
\begin{equation}\label{fbd}
(\eta_{\mu\nu}-F_{\mu\nu})\psi^{\nu}\big|_{\partial\Sigma}
=\epsilon(\eta_{\mu\nu}+F_{\mu\nu}){\bar\psi}^{\nu}\big|_{\partial\Sigma}\,,
\end{equation}
where $\epsilon=\pm$.
Without the background gauge field, the following operator which represents
the tachyon field $T(x)=\sqrt2\lambda\cos(x/\sqrt2)$,
\begin{equation}\label{TX}
-i\sqrt2\lambda\psi \sin(X/\sqrt2)\otimes\sigma_1
\end{equation}
is known to be exactly marginal. Here we have assigned the Chan-Paton
factor $\sigma_1$ and the relevant boundary state is
\begin{equation}\label{bstate2}
|\mathcal{B},\epsilon\rangle_{X,\psi} = \sum_{j = 0,1,\ldots} \sum_{m=-j}^j D_{m,-m}^j(R)
|j;m,m,\epsilon\rangle\rangle,
\end{equation}
where $ | j;m,m,\epsilon \rangle\rangle $ is the
super-Virasoro-Ishibashi state built over the primary state
$ |j;m,m,\epsilon\rangle$ and $\epsilon=\pm$ correspond to
the two different boundary conditions for the fermions in (\ref{fbd}).

Taking the inverse Wick rotation, the deformation (\ref{TX})
describes the rolling tachyon in superstring theory \cite{Sen:2002in}.
The boundary state for the D$p$-brane with this interaction
is given by
\begin{equation}
|\mathcal{B}\rangle=|\mathcal{B},+ \rangle-|\mathcal{B},-\rangle,
\end{equation}
where
\begin{equation}
|\mathcal{B},\epsilon \rangle = |\mathcal{B},\epsilon \rangle_{X^0,\psi^0}
  \otimes_{\mu=1}^p | N,\epsilon \rangle_{X^\mu,\psi^\mu}
  \otimes_{i=p+1}^9 | D,\epsilon \rangle_{X^i,\psi^i}
 \otimes |\text{ghost},\epsilon \rangle.
\end{equation}
Here $|\mathcal{B},\epsilon\rangle_{X^0,\psi^0}$ is the boundary state
(\ref{bstate2}) in the Wick-rotated variables $X = -iX^0$, $\psi=-i\psi^0$, and
$\bar\psi=-i\bar\psi^0$, and
the spatial and ghost parts are usual ones, which are
respectively given by
\begin{align}
|N,\epsilon \rangle_{X^\mu,\psi^\mu}
  &= \exp\left[ -\sum_{n=1}^\infty \frac1n \alpha_{-n}^\mu \bar\alpha_{-n}^\mu
-i\epsilon\sum_{r=1/2}^\infty\psi_{-r}^\mu\bar\psi_{-r}^\mu\right]|0\rangle,
\nonumber \\
|D,\epsilon \rangle_{X^i,\psi^i}
   &= \exp\left[ \sum_{n=1}^\infty \frac1n \alpha_{-n}^i \bar\alpha_{-n}^i
+i\epsilon\sum_{r=1/2}^\infty\psi_{-r}^i\bar\psi_{-r}^i\right]|0\rangle,
 \nonumber \\
|\text{ghost},\epsilon \rangle
   &= \exp\left[ -\sum_{n=1}^\infty (\bar b_{-n} c_{-n} + b_{-n} \bar c_{-n})
+i\epsilon\sum_{r=1/2}^\infty (\beta_{-r}\bar\gamma_{-r}-
\bar\beta_{-r}\gamma_{-r})\right] |\Omega\rangle, \nonumber \\
 |\Omega\rangle&=(c_0+\bar c_0)c_1\bar c_1 e^{-\phi(0)-\bar\phi(0)}|0\rangle.
\end{align}

Now we consider the marginality condition of the tachyon vertex
operator with momentum $k$ in the presence of the constant
electromagnetic field. The vertex operator in the zero-picture is given by
\begin{equation}
-\sqrt2k\cdot\Psi e^{ik\cdot X(t)}.
\end{equation}
Since the fermion $\Psi$ has conformal weight $1/2$, the marginality
condition becomes
\begin{equation}
G^{\mu\nu}k_\mu k_\nu=\frac12.
\end{equation}
If we consider the operators which depend only on $X^1$,
this condition reduces to
\begin{equation}\label{sk1}
k_1^2 =\frac1{2G^{11}}
  = \frac{{\cal Y}_{p}}{2C^{11}}.
\end{equation}

In the absence of the electromagnetic field, $C^{11} = {\cal Y}_p = -1$
and hence $k_1^2 = 1$ and $1/2$, respectively, for the bosonic and superstring
case, as it should be. As the electromagnetic field is turned on
$C^{11}$ and ${\cal Y}_p$ change and so does $k_1^2$. It has a
positive value as long as both ${\cal Y}_p$ and $C^{11}$ are negative.
In (\ref{k12}), this is the case when $E_2^2<1$ with an appropriate $B$
to keep ${\cal Y}_2$ negative.
The marginal tachyon vertex operator is then of the trigonometric type:
$T(X) = \lambda \cos \left(k_{1} X\right)$  or
$\lambda \sin \left(k_{1} X\right)$.
If we examine the corresponding energy-momentum tensor and the R-R coupling,
the resulting configuration with $\lambda = 1/2$ for pure tachyon
case (${\cal Y}_{p}=C^{11}=-1$)
is interpreted as an array of D-branes for the bosonic case
or an array of D($p-1$)$\bar {\rm D}(p-1)$ for
superstrings~\cite{Sen:1998tt,Sen:1998ex}.
It was also discussed in the presence of electric field
(${\cal Y}_{1}=-1+E^2,$ $C^{11}=-1$)~\cite{Sen:2003bc,Kim:2003in,Kim:2003ma}.

On the other hand,
if the electromagnetic field is sufficiently strong, ${\cal Y}_p$ and/or
$C^{11}$ can be flipped to be positive. From the physical ground,
the determinant ${\cal Y}_p$ should be
negative and then the question is whether $C^{11}$ can become
positive while keeping ${\cal Y}_p$ negative.
It turns out that this is possible when $p\ge 2$ as we see in (\ref{k12}).
Note that when $E_2^2 >1$, $C^{11}$ becomes positive while ${\cal Y}_p$ remains
negative as long as the magnetic field $B$ is sufficiently strong.

Then the tachyon profile becomes of the hyperbolic type
$T(X) = e^{\pm\kappa X}$ where
\begin{equation}\label{kappa}
\kappa\equiv i k_1
\,=\left\{
  \begin{array}{ll}
    \sqrt{\frac{-{\cal Y}_{p}}{C^{11}}} &
    \mbox{for}\,\,\mbox{bosonic}\,\,\mbox{string}, \\
    \sqrt{\frac{-{\cal Y}_{p}}{2 C^{11}}} &
    \mbox{for}\,\,\mbox{superstrings}. \\
  \end{array}
\right.
\end{equation}
This is in contrast with the rolling tachyons in which
$k_0^2 = {\cal Y}_{p}/C^{00}$ is always negative.
Actually $C^{00}$ is positive irrespective of the values of
constant electromagnetic field and the dimension $p$ of D-brane.
Therefore turning on the electromagnetic field does not give rise
to a new type of deformation for the case of rolling tachyons.

Without loss of the generality, the tachyon profile
may be classified into the following three cases depending on the asymptotic
behaviors,
\begin{equation}\label{lv}
\frac{T(X)}{\alpha}=\begin{cases}
\mbox{(i)} & \lambda \sinh(\kappa X) \,,\\
\mbox{(ii)} & \lambda \exp(\pm \kappa X)\,,\\
\mbox{(iii)} & \lambda \cosh(\kappa X) \,,
\end{cases}
\end{equation}
where $\alpha=1$ for bosonic string and $\alpha=\sqrt2$ for superstrings.
Note that the coordinate $X$
is a spatial direction along the D-brane. Nevertheless the form of
the operator looks like that of rolling tachyons thanks to the
strong electromagnetic field. This deformation is however entirely physical
and can be obtained through a chain of maps involving T-duality,
Lorentz boost, and rotation~\cite{Rey:2003xs} as discussed in subsection
\ref{subsection3-1} where the corresponding boundary state is constructed.

One comment is in order. For bosonic string, the boundary term
is the same as the tachyon profile (\ref{lv}).
For superstrings, tachyon vertex operators corresponding to (\ref{lv})
in the $-1$-picture are of the form $e^{-\phi}T(X)$.
Since the picture number of the boundary term should be
zero, each boundary term corresponding to the tachyon vertex
(\ref{lv}) is obtained by picture-changing from $-1$ to 0-picture,
\begin{equation}\label{bdytm}
\begin{cases}
\mbox{(i)} & 2i\lambda\kappa\Psi\cosh(\kappa X)\otimes\sigma_1 \,,\\
\mbox{(ii)}&\pm2i\lambda\kappa\Psi\exp(\pm \kappa X)\otimes\sigma_1\,,\\
\mbox{(iii)} & 2i\lambda\kappa\Psi\sinh(\kappa X)\otimes\sigma_1 \,,
\end{cases}
\end{equation}
where the Chan-Paton factor $\sigma_1$ is necessary to describe
GSO-odd states.

The tachyon profiles (\ref{lv}) can also be obtained
in the framework of OSFT.
Let us consider the linearized equations of motion in OSFT
ignoring the interaction among various fields except the coupling to
constant electromagnetic field.
Near the perturbative string vacuum, the tachyonic degree due to
the instability of an unstable D-brane can be described
by a real scalar field $T$ and its action in the
presence of the constant electromagnetic field is expressed in terms of
an open string metric $G^{\mu\nu}$ and a noncommutativity parameter
$\theta^{\mu\nu}$ in (\ref{op}),
\begin{equation}\label{lac}
S_{{\rm L}}
=\int d^{p+1}x\sqrt{-G}\left(-\frac{G^{\mu\nu}}{2}\partial_{\mu}T
\ast\partial_{\nu}T-\frac{m^{2}}{2}T\ast T\right),
\end{equation}
where $G=\det G_{\mu\nu}$ and $\ast$ denotes star product between
the tachyon fields. $m^{2}<0$ is the square of the tachyon mass which is equal
to $-1$ for bosonic string theory and $-1/2$ for superstring theory in
our convention.
Since the background electromagnetic field is constant on the flat
D-brane with the metric $\eta_{\mu\nu}$,
both $G^{\mu\nu}$ and $\theta^{\mu\nu}$ in (\ref{op}) are also constant.
In addition,
every star product in the action (\ref{lac}), quadratic in the tachyon,
can be replaced by an ordinary product and the equation of
motion for the tachyon field becomes
\begin{equation}\label{le1}
G^{\mu\nu}\partial_{\mu}\partial_{\nu}T=m^{2}T.
\end{equation}

For the static kink configurations of codimension-one objects, we assume
$T=T(x), (x=x^{1})$, and then the equation of motion (\ref{le1}) reduces to
\begin{equation}\label{le2}
-C^{11}T^{\prime\prime}=-{\cal Y}_{p}m^{2}T,
\end{equation}
where the prime ${}^{\prime}$ denotes differentiation of $x$.
To keep the role of spacetime variables, the determinant ${\cal Y}_{p}$
should be
nonpositive and, to obtain nontrivial configurations, $C^{11}$ should
be nonvanishing.

As discussed in the previous subsection, the types of the solution of
(\ref{le2}) depend on $k_1^2 = {\cal Y}_p/C^{11}|m^2|$.
When $C^{11}$ is positive, the solution is given by (\ref{lv}).
In the absence of the electromagnetic field, $-C^{11}=1$ and
although the obtained tachyon configurations (\ref{lv}) are static solutions of
the linearized tachyon equation with the coupling of constant electromagnetic
field, the linear tachyon system is obtained through a consistent truncation
of full open string
field equations restricting the fields to a universal subspace and then
the obtained solutions (\ref{lv}) are expected to be solutions of
full open string equations~\cite{Sen:2002nu,Sen:2002in,Mukhopadhyay:2002en}.

\section{Construction of Boundary States for New Codimension-one Objects}
\label{section3}

In this section we construct the boundary state for the hyperbolic type of
marginal deformation (\ref{lv}) along a spatial direction in the presence
of a strong constant electromagnetic field.
For definiteness we concentrate on a D25-brane in bosonic string theory.
Then the case of superstring theory will briefly be discussed.
The generalization to lower-dimensional D-branes is straightforward.

\subsection{T-duality approach}
\label{subsection3-1}

In the following we shall construct the corresponding boundary state through
a chain of transformations starting from a well-established configuration
which turns out to be the rolling tachyon in the presence of the constant
electric field. The order of the transformations is as follows.
We begin with a D25-brane where
a constant electric field is turned on along the $y^1$-direction
and all the other excitations are set to zero except the rolling tachyon.
We compactify the $y^2$-direction on a circle and T-dualize the D25-brane
to a D24-brane. Then we boost it twice: first along the $y^1$-direction
and then along the $y^2$-direction. Subsequently we rotate it in the
$y^1y^2$-plane. Finally we T-dualize it back along the $y^2$-direction.
The resulting configuration will be a D25-brane with the deformation given
in (\ref{lv}).

\begin{figure}[ht]
\begin{center}
\scalebox{1.1}[1.1]{\includegraphics{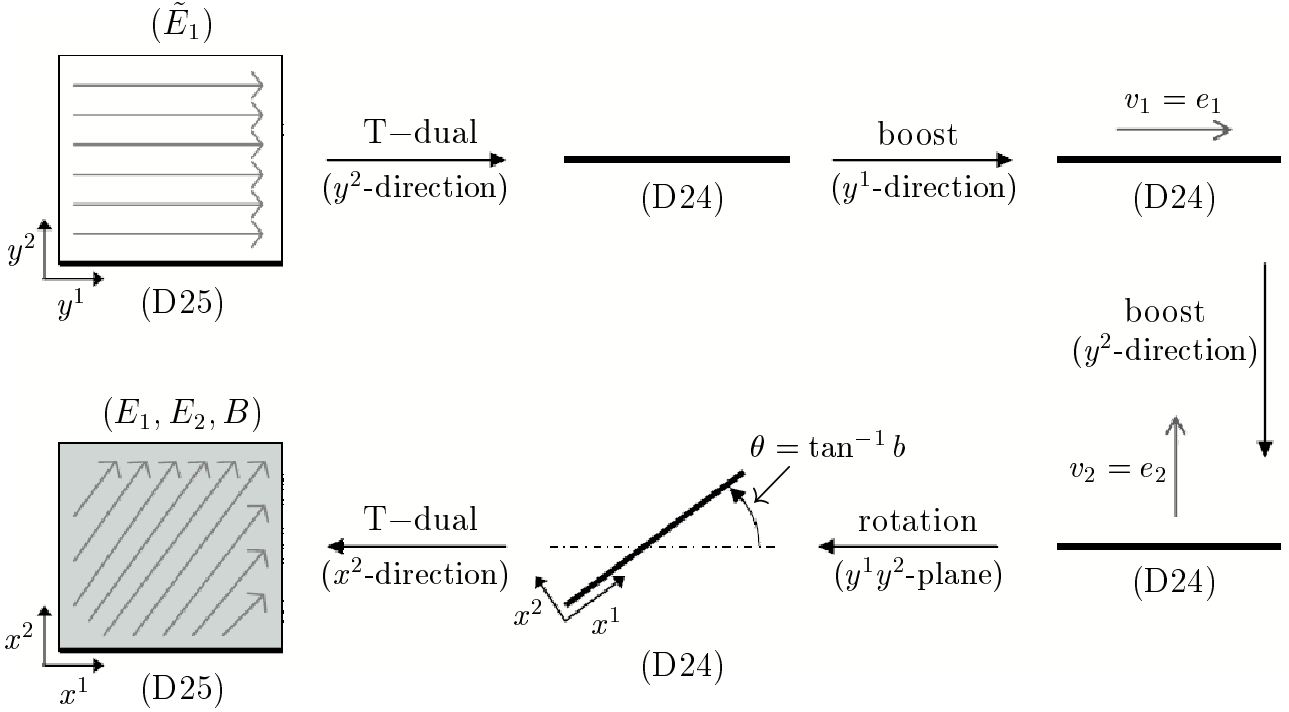}}
\par
\vskip-2.0cm{}
\end{center}
\caption{
{\small The arrows in D25-branes represent electric fields.
The gray background in the second D25-brane represents a nonvanishing
magnetic field.}}
\label{fig1}
\end{figure}

Let us first consider a flat D25-brane ($p=25$) with rolling tachyon
in a constant electric field. We denote the worldsheet fields of the brane
by $Y^\mu$ and assume that the constant electric field denoted by
$\tilde E_1$ is along the $y^1$-direction.
The rolling tachyon is then described
by the exactly marginal boundary operator\footnote{We display only the
    cosh-type operator for simplicity.}~\cite{Mukhopadhyay:2002en}
\begin{equation} \label{chy}
\lambda \int dt \cosh\left( \sqrt{1-\tilde E_1^2}\, Y^0(t) \right).
\end{equation}
We compactify the $y^2$-direction on a circle of radius $R$ and wrap the
D25-brane on it.

Under the T-dualization of the $y^2$-direction, the right-moving part
of $Y^2$ changes its sign
while the other fields remain unchanged,
\begin{equation}
\mbox{T dual:}\quad Y_R^2(\bar z) \longrightarrow -Y_R^2(\bar z).
\end{equation}
The D25-brane is then turned into an array of D24-branes on the dual circle.
Taking the decompactification limit $R\rightarrow\infty$ we get
a localized D24-brane with a constant electric field $\tilde E_1$
turned on along the $y^1$-direction. The $Y^2$ part of the
boundary state is given by the Dirichlet state,
\begin{equation} \label{dstate}
|D \rangle_{Y^2}
  = \exp\left( \sum_{n=1}^\infty\frac1n\beta_{-n}^2\bar\beta_{-n}^2 \right)
    \delta({\hat y^2}) |0 \rangle\,,
\end{equation}
where $\beta_n^\mu$ and $\bar\beta_n^\mu$ denote the oscillators of $Y^\mu$.

Next, we boost the D24-brane along the $y^1$-direction with a velocity $e_1$
which is followed by another boost along the $y^2$-direction with a velocity
$e_2$. Then we perform a rotation in the $y^1y^2$-plane by an angle
$\tan^{-1}b$.
Then $Y^0,Y^1,Y^2$ are mapped as
\begin{equation}
\begin{pmatrix} Y^0 \\ Y^1 \\ Y^2 \end{pmatrix}
\longrightarrow
\begin{pmatrix} X^0 \\ X^1 \\ X^2 \end{pmatrix}
 = \Omega\Lambda_2\Lambda_1 \begin{pmatrix} Y^0 \\ Y^1 \\ Y^2 \end{pmatrix}.
\end{equation}
Here the boost transformations, $\Lambda_1$ and $\Lambda_2$, and
the rotation $\Omega$ are
\begin{align}\label{LLO}
\Lambda_1&= \begin{pmatrix}
             \gamma_1     & \gamma_1 e_1 & 0 \\
             \gamma_1 e_1 & \gamma_1     & 0 \\
             0            & 0            & 1
             \end{pmatrix},
&\gamma_1 &= \frac{1}{\sqrt{1-e_1^2}}, &&e_1^2 <1, \nonumber \\
\Lambda_2 &= \begin{pmatrix}
             \gamma_2     & 0 & \gamma_2 e_2 \\
             0            & 1 & 0            \\
             \gamma_2 e_2 & 0 & \gamma_2
             \end{pmatrix},
&\gamma_2 &= \frac{1}{\sqrt{1-e_2^2}}, &&e_2^2 <1, \nonumber \\
\Omega &= \begin{pmatrix}
               1 & 0              & 0               \\
               0 & \tilde\gamma   & -\tilde\gamma b \\
               0 & \tilde\gamma b & \tilde\gamma
          \end{pmatrix},
&\tilde\gamma&=\frac1{\sqrt{1+b^2}}, &&0< b^2 <\infty,
\end{align}
and then the resulting transformation matrix is
\begin{equation} \label{OL1L2}
\Omega\Lambda_2\Lambda_1 =
  \begin{pmatrix}
    \gamma_1\gamma_2 & \gamma_1\gamma_2 e_1 & \gamma_2 e_2 \\
    -\gamma_1\gamma_2\tilde\gamma e_2 b + \gamma_1\tilde\gamma e_1 &
    -\gamma_1\gamma_2\tilde\gamma e_1 e_2 b + \gamma_1\tilde\gamma &
    -\gamma_2\tilde\gamma b \\
    \gamma_1\gamma_2\tilde\gamma e_2
    + \gamma_1\tilde\gamma e_1 b&
    \gamma_1\gamma_2\tilde\gamma e_1 e_2 +\gamma_1\tilde\gamma b &
    \gamma_2\tilde\gamma
  \end{pmatrix}.
\end{equation}
Now we choose the rotation parameter $b$ as
\begin{equation} \label{eeb}
b = \frac1{\gamma_2e_1 e_2},
\end{equation}
so that the 11-component of $\Omega\Lambda_2\Lambda_1$ in (\ref{OL1L2})
vanishes.
The delta function $\delta({\hat y^2})$ in the
boundary state (\ref{dstate}) then becomes
\begin{align} \label{delta2}
\delta (\hat y^2)
  &= \delta (-\gamma_2 e_2 \hat x^0 - \gamma_2\tilde\gamma b \hat x^1
             +\gamma_2\tilde\gamma \hat x^2) \nonumber \\
  &= (\gamma_2\tilde\gamma)^{-1}
      \delta ( \hat x^2 - \tilde\gamma^{-1}e_2 \hat x^0 - b\hat x^1 ).
\end{align}
From (\ref{OL1L2})--(\ref{delta2}), we find
that within the boundary state
\begin{equation} \label{ytox}
\hat y^0 = - e_1\gamma_1\tilde\gamma^{-1} \hat x^1, \qquad
\hat y^1 = \frac{e_2 b}{\gamma_1} \,\hat x^0
           + \frac{\gamma_1}{\tilde\gamma}\,\hat x^1.
\end{equation}
Note that the zero mode $\hat y^0$ in the zeroth direction is
transformed to $\hat x^1$ with the help of zero-mode constraint
in (\ref{dstate}) for the localized D24-brane.

As the last step, we compactify the $x^2$-direction on a circle
and T-dualize back along the $x^2$-direction, which changes
$X_R^2(\bar z)$ to $-X_R^2(\bar z)$. Finally we decompactify
the $x^2$-direction. The whole transformation can then be written as
\begin{align} \label{total}
\begin{pmatrix}
    X_L^0 \\ X_L^1 \\ X_L^2 \\
\end{pmatrix}
&=\Omega\Lambda_2\Lambda_1
   \begin{pmatrix}
    Y_L^0 \\ Y_L^1 \\ Y_L^2 \\
   \end{pmatrix}, \nonumber \\[2mm]
   \begin{pmatrix}
    X_R^0 \\ X_R^1 \\ X_R^2 \\
   \end{pmatrix}
&= \begin{pmatrix}
    1 & 0 & 0 \\
    0 & 1 & 0 \\
    0 & 0 & -1 \\
   \end{pmatrix}
   \Omega\Lambda_2\Lambda_1
   \begin{pmatrix}
    1 & 0 & 0 \\
    0 & 1 & 0 \\
    0 & 0 & -1 \\
   \end{pmatrix}
   \begin{pmatrix}
    Y_R^0 \\ Y_R^1 \\ Y_R^2 \\
   \end{pmatrix}
=\Omega^{-1}\Lambda_2^{-1}\Lambda_1
   \begin{pmatrix}
    Y_R^0 \\ Y_R^1 \\ Y_R^2 \\
   \end{pmatrix}.
\end{align}
Moreover, the zero-mode part is transformed to
\begin{equation}  \label{zeromode2}
(\gamma_2\tilde\gamma)^{-1}
      \delta(\hat x^2-\tilde\gamma^{-1}e_2\hat x^0-b\hat x^1)|0\rangle
\longrightarrow (\gamma_2\tilde\gamma)^{-1} | 0 \rangle,
\end{equation}
yielding the Born-Infeld factor $(\gamma_2\tilde\gamma)^{-1}$,
since only the zero-winding sector survives in the decompactification limit.

After the chain of these transformations, we get a D25-brane with constant
electric and magnetic fields deformed by the tachyon field
\begin{equation} \label{ch}
\lambda \cosh\left( \sqrt{1-\tilde E_1^2}\,
e_1\gamma_1\tilde\gamma^{-1} X^1 \right),
\end{equation}
where (\ref{ytox}) has been used.
The resulting electromagnetic fields $E_1,E_2,B$ on the D-brane are
related to the boost and rotation parameters through
\begin{align}\label{eeb2}
\tilde E_1 = \frac{E_1}{\sqrt{1-E_2^2 + B^2}}, \quad
e_1 = \frac{\sqrt{1-E_2^2 + B^2}}{E_2 B}, \quad
e_2 = \frac{E_2}{\sqrt{1 + B^2}}, \quad
b = B.
\end{align}
To see this, and also as a check of our calculation, we
apply the transformations (\ref{total}) and (\ref{zeromode2})
to the Neumann boundary state of a static
D25-brane with a nonvanishing constant electric field
$\tilde E_1=\tilde F_{01} =- \tilde F_{10}$. The nontrivial part is
\begin{equation}
|N;\tilde E^1 \rangle_{Y^{0,1,2}} = N_p\sqrt{-{\tilde {\cal Y}}_{p}}
   \exp\left[ -\sum_{n=1}^\infty \frac1n
             \beta_{-n}^a \bar\tau_{-n\,a}
       \right]|0\rangle,\qquad
(a,b=0,1,2),
\end{equation}
where $N_p$ is an overall constant and ${\tilde {\cal Y}}_{p}$
is~\cite{DiVecchia:1999rh}
\begin{equation}
{\tilde {\cal Y}}_{p} =  \det( \eta_{ab}+\tilde F_{ab} )
             = -1+\tilde E_1^2.
\end{equation}
The oscillators $\bar\tau_{-n}$'s are defined as
\begin{equation} \label{taum}
\bar\tau = \tilde M \bar\beta
\equiv \eta \left( \frac{ \eta - \tilde F}{ \eta + \tilde F} \right)\bar\beta.
\end{equation}
From (\ref{total}), the relevant factor in the exponent becomes
\begin{equation}
\beta_{-n}^T \bar\tau_{-n}
 = \alpha_{-n}^T (\Omega \Lambda_2 \Lambda_1)^{-1\,T} \tilde M\,
             \Lambda_1^{-1} \Lambda_2 \Omega\, \bar\alpha_{-n}.
\end{equation}
Using (\ref{eeb2}), it is straightforward to find that this reduces to
\begin{equation}
 \alpha_{-n}^T  \eta \left( \frac{ \eta - F}{ \eta + F} \right)
\bar\alpha_{-n},
\end{equation}
where $F$ is the field strength tensor with electric field $\bf E$ and
magnetic field $B$. This gives the correct form of the exponent for the
Neumann boundary state with constant electric field $\bf E$ and magnetic
field $B$. Furthermore, the zero mode part becomes
\begin{align} \label{bifactor}
(\gamma_2 \tilde\gamma)^{-1} \sqrt{-{\tilde {\cal Y}}_{p}}
 &= \sqrt{1-{\bf E}^{2}+B^2}  \nonumber \\
&= \sqrt{-\det(\eta_{ab}+F_{ab})} = \sqrt{-{\cal Y}_{p}},
\end{align}
reproducing the Born-Infeld factor with ${\bf E}$ and $B$.
This establishes the
relation (\ref{eeb2}). With the identification (\ref{eeb2}), it is
easy to see that the deformation
(\ref{ch}) precisely takes the form (\ref{lv}) with $X=X^1$.

The relation (\ref{eeb2}) constrains the range of the allowed electromagnetic
fields obtained by the transformation (\ref{total}). Since the initial
electric field $\tilde E_1$ on the D-brane and the boost parameters
$e_1,e_2$ should be less than one, we get
\begin{equation} \label{condition}
-{\cal Y}_{p} = 1-{\bf E}^{2} + B^2 > 0,\qquad E_2^2 > 1.
\end{equation}
The first constraint is nothing but the reality condition of the Born-Infeld
factor (\ref{bifactor}). The second constraint $E_2^2 >1$ is precisely
the condition that $\kappa$ in (\ref{lv}) is real so that the deformation
is of the hyperbolic type as discussed before. We note that these conditions
are obtained from a series of boosts and rotations which are not
connected to the identity transformation since $E_2^2$ is necessarily
larger than one.

Now it is a straightforward matter to write down the
boundary state of D25-brane deformed by the boundary operator (\ref{lv})
in the presence of a constant electromagnetic field with the condition
(\ref{condition}). We just apply the transformations
(\ref{ytox})--(\ref{zeromode2}) and (\ref{eeb2})
to the boundary state of the rolling
tachyon with the electric field $\tilde E_1$ calculated
in \cite{Mukhopadhyay:2002en}. More explicitly, suppose that
the matter part of the boundary state with $\tilde E_1$ is given by
\begin{equation}
|\mathcal{B};\tilde E_1 \rangle_m = N_p\sqrt{-{\tilde {\cal Y}}_{p}}\,
 [ -\tilde B(Y^0(0)) |0 \rangle
 + \beta_{-1}^\mu\bar\beta_{-1}^\nu \tilde A_{\mu\nu}(Y^0(0)) |0 \rangle
 + \cdots].
\end{equation}
Then the matter part of the corresponding boundary state with
the electromagnetic field becomes
\begin{align}
|\mathcal{B};F \rangle_m = &N_p\sqrt{-{\cal Y}_{p}}
\left\{
 -\tilde B
  \left(- e_1\gamma_1 \tilde\gamma^{-1}\sqrt{1-\tilde E_1^2}\,X^1(0)\right)
 | 0 \rangle \right. \nonumber \\
& +\left.\alpha_{-1}^\mu \bar\alpha_{-1}^\nu
\left[(\Lambda_1^{-1}\Lambda_2^{-1}\Omega^{-1})^T
\tilde A
  \left(- e_1\gamma_1\tilde\gamma^{-1}\sqrt{1-\tilde E_1^2}\, X^1(0)\right)
(\Lambda_1^{-1}\Lambda_2\Omega)\right]_{\mu\nu}| 0 \rangle
+ \cdots \right\},
\end{align}
where (\ref{bifactor}) was used.
Using the explicit form of $\tilde A_{\mu\nu}(x)$ and $\tilde B(x)$ given in
\cite{Mukhopadhyay:2002en,Rey:2003xs} we find that
\begin{equation} \label{bt}
|\mathcal{B};F \rangle_m = N_p\sqrt{-{\cal Y}_{p}}\,
 [ -B(X^1(0)) |0 \rangle
 + \alpha_{-1}^\mu\bar\alpha_{-1}^\nu A_{\mu\nu}(X^1(0)) |0 \rangle + \cdots],
\end{equation}
where
\begin{align}
B(x^1) =& -f(-\kappa x^1), \nonumber \\
A^{\mu\nu} (x^1)=& -2\left[-\frac12 \eta^{\mu\nu}
+ G^{\mu\nu}+\theta^{\mu\nu}
-\frac1{G^{11}} \left(G^{\mu 1} + \theta^{\mu 1}\right)
\left(G^{\nu 1} - \theta^{\nu 1}\right)\right] f(-\kappa x^1)
\nonumber \\
&-\frac1{G^{11}} \left(G^{\mu 1} + \theta^{\mu 1}\right)
\left(G^{\nu 1} - \theta^{\nu 1}\right) (\hat\lambda+ 1).\label{Amunu}
\end{align}
Here $G^{\mu\nu}$ and $\theta^{\mu\nu}$ are defined in (\ref{op}).
The function $f(x)$ and $\hat\lambda$ are given in the following form.
For the sinh-type profile ((i) in (\ref{lv})), they are\footnote{This result
can be trusted only for $|\sinh(\pi\lambda)|<1$ \cite{Sen:2004nf}.}
\begin{equation}\label{fs}
f(x)=\frac{1}{1+e^{x}\sinh \pi\lambda}
+\frac{1}{1-e^{-x}\sinh \pi\lambda}-1 ,
\qquad \hat\lambda=\cosh(2\pi\lambda).
\end{equation}
For the exponential type profile ((ii) in (\ref{lv})),
\begin{equation}\label{f0}
f(x)=\frac{1}{1+2\pi\lambda\,e^{-x}},\qquad \hat\lambda=1.
\end{equation}
For the cosh-type profile ((iii) in (\ref{lv})),
\begin{equation}\label{fc}
f(x)=\frac{1}{1+e^{x}\sin \pi\lambda}
+\frac{1}{1+e^{-x}\sin \pi\lambda}-1,
\qquad \hat\lambda=\cos(2\pi\lambda).
\end{equation}

As mentioned in \cite{Rey:2003xs}, the generalization to superstrings
is straightforward. Again we start with the boundary state with
the electric field $\tilde E_1$ in superstring theory,
\begin{equation}
|\mathcal{B};\tilde E_1,\epsilon \rangle_m = N_p\sqrt{-{\tilde {\cal Y}}_{p}}\,
 [ -\tilde B(Y^0(0)) |0 \rangle
 +i\epsilon \chi_{-1/2}^\mu\bar\chi_{-1/2}^\nu \tilde A_{\mu\nu}(Y^0(0))
|0 \rangle + \cdots],
\end{equation}
where $\tilde A_{\mu\nu}$  and $\tilde B$ are the same as in the bosonic
case except for the form of the function $f(x)$, and $\chi$ and $\bar\chi$
are the oscillators of the fermionic partner of $Y$.
After the chain of the Lorentz transformation and T-duality,
we obtain
\begin{equation}
|\mathcal{B};F,\epsilon \rangle_m = N_p\sqrt{-{\cal Y}_{p}}\,
 [ -B(X^1(0)) |0 \rangle
+i\epsilon \psi_{-1/2}^\mu\bar\psi_{-1/2}^\nu A_{\mu\nu}(X^1(0))
|0 \rangle + \cdots],\label{sbt}
\end{equation}
where $A_{\mu\nu}$ and $B$ are already given in (\ref{Amunu}).

For the sinh-type profile ((i) in (\ref{lv})),
the function $f(x)$ and $\hat\lambda$ have the form\footnote{As in the bosonic
case, this is valid only for $|\sinh(\pi\lambda)|<1$.}
\begin{equation}\label{sfs}
f(x)=\frac{1}{1+\sinh^2(\pi\lambda)e^{2x}}
+\frac{1}{1+\sinh^2(\pi\lambda)e^{-2x}}-1,
\qquad \hat\lambda=\cosh(2\pi\lambda),
\end{equation}
for the exponential type profile ((ii) in (\ref{lv})), they are
\begin{equation}\label{sf0}
f(x)=\frac{1}{1+4\pi^{2}\lambda^{2}e^{\mp2x}},
\qquad \hat\lambda=1,
\end{equation}
and, for the cosh-type profile ((iii) in (\ref{lv})),
\begin{equation}\label{sfc}
f(x)=\frac{1}{1+\sin^2(\pi\lambda)e^{2x}}
+\frac{1}{1+\sin^2(\pi\lambda)e^{-2x}}-1,
\qquad \hat\lambda=\cos(2\pi\lambda).
\end{equation}

The above form of the boundary state is enough \cite{Sen:2002in}
to obtain the energy-momentum tensor and the current density of fundamental
strings in section \ref{section4}.
The closed form of the boundary state can also be obtained from the
boundary state with $\tilde E_1$ in \cite{Mukhopadhyay:2002en} by
applying the transformations as mentioned above. Instead of doing
this, in the next subsection, we shall obtain the closed
form of the boundary state by directly dealing with the
deformation (\ref{lv}).

\subsection{Boundary conformal field theory}
\label{subsection3-2}

In this subsection we shall construct the corresponding boundary state
following the method used in \cite{Mukhopadhyay:2002en}.
This calculation will also serve as a consistency check of the previous
approach using T-duality.
Then we generalize the result to superstring case.
Although we concentrate on the operator in (iii) of (\ref{lv}),
it is straightforward to generalize the result to
other tachyon vertices in (i) and (ii) of (\ref{lv}).

In the Wick-rotated theory obtained by the replacement $X^0 \rightarrow
-iX^0$, we introduce the vielbeins of the open and closed string metrics
and the corresponding local coordinates as follows\footnote{
We put bars ($\; \bar{} \; $) to denote quantities in Wick rotated theory.}
\begin{align}
\bar G_{\mu\nu} &= V_\mu^a V_\nu^a = (V^T V)_{\mu\nu},
       && \delta_{\mu\nu} = v_\mu^a v_\nu^a = (v^T v)_{\mu\nu}, \nonumber \\
Z^a &= V_\mu^a X^\mu, && W^a = v_\mu^a X^\mu,
\end{align}
where $a,\mu = 0,1,\ldots,25$. Explicitly,\footnote{For simplicity,
we will not explicitly display the components
in the trivial directions $a,\mu=3,4,\ldots,25$.}
\begin{equation}
\bar G = 1-\bar F^2 = \begin{pmatrix}
1+\bar E_1^2 + \bar E_2^2  & B \bar E_2          & -B \bar E_1 \\
B \bar E_2                 & 1+ B^2 + \bar E_1^2 & \bar E_1 \bar E_2 \\
-B \bar E_1                & \bar E_1 \bar E_2   & 1+ B^2 +\bar E_2^2
\end{pmatrix}
\end{equation}
and we choose
\begin{align}
V &= \begin{pmatrix}
0 & -\sqrt{\frac{{\bar {\cal Y}}_{2}}{1 + \bar E_2^2}}    & 0 \\
\sqrt{\frac{{\bar {\cal Y}}_{2} (1 + \bar E_2^2)}{1+B^2 + \bar E_2^2}}  &
  B \bar E_2 \sqrt{\frac{{\bar {\cal Y}}_{2}}{(1 + \bar E_2^2)
(1+B^2 + \bar E_2^2)}} & 0 \\
-\frac{B \bar E_1}{\sqrt{1+B^2 + \bar E_2^2}}  &
    \frac{\bar E_1 \bar E_2}{\sqrt{1+B^2 + \bar E_2^2}}  &
    \sqrt{1+B^2 + \bar E_2^2}
\end{pmatrix},\nonumber \\
v &= V (1-\bar F)^{-1},\label{vielbein}
\end{align}
where ${\bar {\cal Y}}_{2} = \det(1+\bar F) = 1+{\bar {\bf E}}^{2}+ B^2$
and $v$ chosen in this way is indeed an orthogonal matrix. The vielbein $V$
is chosen in such a way that boundary operator $J_{Z^0}^1(t) = \cos (Z^0(t))$
becomes (\ref{lv}) with $X=X^1$ after inverse Wick rotation, i.e.,
\begin{align} \label{cosz}
\lambda \int dt J_{Z^0}^1(t)
 &= \lambda \int dt \cos (Z^0(t)) \nonumber \\
 &= \lambda \int dt \cos({\bar \kappa} X^1), \qquad
      {\bar \kappa} = \sqrt{\frac{{\bar {\cal Y}}_{2}}{1 + \bar E_2^2}},
\end{align}
which describes an exactly marginal deformation.
Also the components in the second row of $V$ are chosen to reproduce
the relation (\ref{ytox}). In this way the vielbein $V$
has a clear interpretation in the T-duality approach.
The relation between the two local frames is given by
\begin{equation} \label{zy}
Z = (1 - \hat F)W, \qquad \hat F = v \bar F v^T.
\end{equation}

Now we compactify the $W^0$ coordinate on a circle of unit radius,
\begin{equation} \label{ycpt}
W^0 \sim W^0 + 2\pi.
\end{equation}
Since the closed string metric is identity in the $W^a$ coordinate system,
(\ref{ycpt}) implies that the closed string theory now has enhanced
${\rm SU}(2)_L \times {\rm SU}(2)_R$
gauge symmetry which can be used to organize the
boundary state as in \cite{Callan:1993mw,Recknagel:1998ih}. The
left-moving currents are given by
\begin{equation}
J_{W_L^0}^1 = \cos(2W_L^0),\qquad
J_{W_L^0}^2 = \sin(2W_L^0),\qquad
J_{W_L^0}^3 = i\partial W_L^0,
\end{equation}
which are all well-defined operators since $W^0$ is compactified on a circle
of self-dual radius. From (\ref{zy}), (\ref{ycpt}) implies that
\begin{equation}
(Z^0,Z^1,Z^2) \sim
(Z^0 + 2\pi,\ Z^1 - 2\pi \hat F_{10} ,\ Z^2 - 2\pi \hat F_{20} ),
\end{equation}
under which the operator $\cos(Z^0)$ is manifestly invariant. Therefore,
with this compactification, we can obtain the deformed boundary state
$|{\cal B}; \bar F; \lambda \rangle$ starting from the unperturbed
($\lambda=0$) boundary state $|{\cal B}; \bar F\rangle$
which is constructed below.

We shall now construct the boundary state $|{\cal B}; \bar F\rangle$
for an Euclidean D25-brane ($p=25$) with an electromagnetic field $\hat F_{\mu\nu}$
turned on, and $W^0$ compactified on a circle of unit radius. From
the boundary condition (\ref{bnd1}) the closed string overlap condition
reads
\begin{equation}
(\partial_\tau X + i\bar F \partial_\sigma X)|_{\tau=0} = 0.
\end{equation}
In terms of the oscillators $\sigma_n$'s and $\bar\sigma_n$'s of the
coordinates $W$, this becomes
\begin{equation} \label{overlap}
(\sigma_n + \hat M \bar \sigma_{-n}) |{\cal B};\bar F \rangle = 0,
 \qquad  n \in \mathbb{Z},
\end{equation}
where
\begin{align} \label{tm}
\hat M = \frac{1-\hat F}{1+\hat F}
         =v \frac{1-\bar F}{1+\bar F} v^T \equiv v \bar M v^T.
\end{align}
Then we get
\begin{equation} \label{bdry0}
|{\cal B}; \bar F\rangle
 = N_p\sqrt{{\bar {\cal Y}}_p} \exp\left[ -\sum_{n=1}^\infty
                         \frac1n \sigma_{-n}^T \hat M \bar\sigma_{-n}
                  \right]| {\cal B};\bar F\rangle_0
   \otimes |\mbox{ghost}\rangle,
\end{equation}
where $| {\cal B};\bar F\rangle_0$ is the zero mode part of the boundary
state.

To construct the zero mode part, we first compactify the coordinates $W^1$
and $W^2$ on circles with radii $R^{(1)}$ and $R^{(2)}$, respectively. We shall
take the decompactification limit later to get  the desired result.
Let $n^{(a)}, m^{(a)} \in \mathbb{Z}$ ($a=0,1,2$) be the momentum and
winding numbers respectively. Then the overlap condition (\ref{overlap})
for the zero modes reads
\begin{equation}
\left( \frac{n^{(a)}}{R^{(a)}} + m^{(a)} R^{(a)} \right)
+ \hat M^a_{~~b}
\left( \frac{n^{(a)}}{R^{(a)}} - m^{(a)} R^{(a)} \right) = 0,
\end{equation}
with $R^{(0)}= 1$. With (\ref{tm}), this reduces to
\begin{equation}
n^{(a)} = - R^{(a)} \hat F_{ab} R^{(b)} m^{(b)}\qquad
(\mbox{no sum over $a$}).
\end{equation}
Thus $n^{(a)}$'s are not independent and the sum can be taken over
$m^{(a)}$'s only. We get
\begin{equation} \label{zero3}
| {\cal B};\bar F\rangle_0 = \sum_{m^{(a)}\in \mathbb{Z}}
      \exp\left[ \sum_{a=0}^2(i\sigma_0^a w_L^a
+i\bar\sigma_0^a w_R^a) \right]
 |0 \rangle,
\end{equation}
where
\begin{align}
\sigma_0^1
 &= -\hat F_{01}R^{(1)} m^{(1)} - \hat F_{02}R^{(2)} m^{(2)} + m^{(0)},
                                  \nonumber \\
\sigma_0^2
 &= - \hat F_{10} m^{(0)} - \hat F_{12}R^{(2)} m^{(2)} + m^{(1)}R^{(1)},
                                  \nonumber \\
\sigma_0^3
 &= - \hat F_{20} m^{(0)} - \hat F_{21}R^{(1)} m^{(1)} + m^{(2)}R^{(2)},
\end{align}
and we replace $m^{(a)}$ in the last term
of each line by $-m^{(a)}$ to get $\bar\sigma_0^a$ in this
expression. The decompactification limit $R^{(1)}, R^{(2)} \rightarrow\infty$
is obtained by keeping only the $m^{(1)}=m^{(2)}=0$ terms in (\ref{zero3}).
This gives
\begin{equation}
| {\cal B};\bar F\rangle_0 = \sum_{m\in \mathbb{Z}/2}
                  \exp\left[ -2im (w_L^0 - w_R^0)
                     + 2im(\hat F_{10} w^1 + \hat F_{20} w^2) \right]
                 |0 \rangle.
\end{equation}
Inserting the zero mode part into (\ref{bdry0}), we obtain
\begin{align} \label{bdry1}
|{\cal B}; \bar F\rangle
 = &N_p\sqrt{{\bar {\cal Y}}_{p}} \exp\left[ -\sum_{n=1}^\infty
                         \frac1n \sigma_{-n}^a \bar\tau_{-n}^a
                  \right]
  \sum_{m\in \mathbb{Z}/2}
                  \exp\left[ -2im (w_L^0 - w_R^0)
                     + 2im(\hat F_{10} w^1 + \hat F_{20} w^2) \right]
       |0 \rangle \nonumber \\
  &\otimes_{i=3}^{25} |N \rangle_{X^i} \otimes |\mbox{ghost}\rangle,
\end{align}
where the oscillators $\bar\tau_n$ are defined by
\begin{equation} \label{taun}
\bar\tau_n = \hat M \bar\sigma_n = v^{-1}\bar\alpha_n,\qquad n\neq 0,
\end{equation}
with $\bar\alpha_n$ being the oscillators of $X$. Using the Virasoro-Ishibashi
state $|j;m,m\rangle\rangle$ of (\ref{bstate}), this can be written as
\begin{align}
|{\cal B}; \bar F\rangle
= &N_p\sqrt{{\bar {\cal Y}}_p} \sum_{j,m} |j;-m,m\rangle\rangle_{\bar\tau}^{(0)}
   \otimes \exp\left[ - \sum_{n=1}^\infty\frac1n (\sigma_{-n}^1 \bar\tau_{-n}^1
                               + \sigma_{-n}^2 \bar\tau_{-n}^2)
                     + 2im(\hat F_{10} w^1 + \hat F_{20} w^2) \right]
       |0 \rangle \nonumber \\
  &\otimes_{i=3}^{25} |N \rangle_{X^i} \otimes |\mbox{ghost}\rangle,
\end{align}
where $|j;-m,m\rangle\rangle_{\bar\tau}^{(0)}$ is the state obtained
by replacing the $\bar\sigma$ oscillators by the corresponding
$\bar\tau$ oscillators on the right-part of the state appearing
in the expansion of $ |j;-m,m\rangle\rangle^{(0)}$.

Now turn on the deformation (\ref{cosz}). Using the boundary condition,
we see that $Z^0(t) = 2W_L^0(t)$ on the boundary and hence,
on the boundary,
\begin{equation}
\cos(Z^0(t)) = \cos(2W_L^0(t)) = J_{W_L^0}^1(t).
\end{equation}
Then with the ${\rm SU}(2)_L$ charge $Q_{W_L^0}^1$ defined by
\begin{equation}
Q_{W_L^0}^i = \oint \frac{du}{2\pi i} J_{W_L^0}^i(u),
\end{equation}
the boundary state $|{\cal B};\bar F,\lambda \rangle$
in the presence of the boundary deformation (\ref{cosz}) is obtained as
\cite{Recknagel:1998ih}
\begin{align}
|{\cal B};\bar F,\lambda \rangle
 &= \exp( -2\pi i \tilde\lambda Q_{W_L^0}^1 ) |{\cal B};\bar F \rangle
  \nonumber \\
 &\hspace{-13mm}
= N_p\sqrt{{\bar {\cal Y}}_p}
    \sum_{j,m,m'} D_{m',-m}^j |j;m',m\rangle\rangle_{\bar\tau}^{(0)}
\nonumber \\
   &\hspace{-8mm}
\otimes \exp\left[ - \sum_{n=1}^\infty\frac1n (\sigma_{-n}^1 \bar\tau_{-n}^1
                               + \sigma_{-n}^2 \bar\tau_{-n}^2)
                     + 2im(\hat F_{10} w^1 + \hat F_{20} w^2) \right]
       |0 \rangle
\otimes_{i=3}^{25} |N \rangle_{X^i} \otimes |\mbox{ghost}\rangle,
\end{align}
where $D_{m',m}^j $ is the spin $j$ representation of
$\exp( -2\pi i \lambda Q_{W_L^0}^1 )$.

As the final step, we can take the decompactification limit by
removing all the winding sector states. In this limit only the state
with $m'=m$ survives \cite{Callan:1993mw} and we get the desired
boundary state
\begin{align}
\lefteqn{|{\cal B};\bar F,\lambda \rangle}\nonumber\\
&=N_p\sqrt{{\bar {\cal Y}}_{p}}
   \sum_{j,m} D_{m,-m}^j |j;m,m\rangle\rangle_{\bar\tau}^{(0)}
   \otimes \exp\left[ -\sum_{n=1}^\infty\frac1n (\sigma_{-n}^1 \bar\tau_{-n}^1
                           + \sigma_{-n}^2 \bar\tau_{-n}^2)
                     + 2im(\hat F_{10} w^1 + \hat F_{20} w^2) \right]
       |0 \rangle \nonumber \\
  &\hspace{6mm}
\otimes_{i=3}^{25} |N \rangle_{X^i} \otimes |\mbox{ghost}\rangle.
\label{hestate}
\end{align}
From (\ref{zy}) and (\ref{taun}) it is readily seen that,
after the inverse Wick rotation, this precisely reproduces the state
(\ref{bt}) we have obtained in subsection~\ref{subsection3-1}.
This also verifies the consistency of the whole calculations performed
in this subsection.

Based on the obtained result for bosonic string, we discuss
the case of superstrings as in~\cite{Mukhopadhyay:2002en,Sen:2002in}.
With the choice of the vielbein $V$ (\ref{vielbein}), the boundary operator
$-i\sqrt2\lambda\Psi_Z^0 \sin(Z^0/\sqrt2)\otimes\sigma_1$ becomes the
boundary operator (iii) in (\ref{bdytm}) after the inverse Wick rotation, i.e.,
\begin{equation} \label{coszs}
-i\sqrt2\lambda \int dt \Psi^0_{Z}\sin \left(\frac{Z^0(t)}{\sqrt2}\right)
\otimes\sigma_1
=-2i\lambda \int dt {\bar \kappa} \Psi^1 \sin\left(\bar\kappa
 X^1\right)\otimes\sigma_1,
\end{equation}
where $\Psi_Z=V\Psi$ and
$\bar\kappa=\sqrt{\bar{\cal Y}_{2}/2(1 + \bar E_2^2)}$.

Similar to the bosonic case,
we compactify $W^0$ on a circle of radius $\sqrt2$,
\begin{equation}
 W^0 \sim W^0+2\sqrt2\pi.
\end{equation}
Then we have an enhanced ${\rm SU}(2)_L\times {\rm SU}(2)_R$ symmetry.
The left-moving ${\rm SU}(2)$ currents are given by
\begin{equation}
J_{W_L^0}^1 =-i\sqrt2 \psi^0_{W}\sin(\sqrt2W_L^0)
\otimes\sigma_1,\quad
J_{W_L^0}^2 =i\sqrt2 \psi^0_{W}\cos(\sqrt2W_L^0)
\otimes\sigma_1,\quad
J_{W_L^0}^3 = i\sqrt2 \,\partial W_L^0,
\end{equation}
where $\psi_W=v\psi$. Then the ${\rm SU}(2)$ charges are defined as
\begin{equation}
Q_{W_L^0}^i = \oint \frac{dz}{2\pi i} J_{W_L^0}^i(z).
\end{equation}

It is known that a GSO-invariant boundary state is given by
the linear combination of the two boundary states corresponding
to the different boundary condition for the fermions,
\begin{equation}
|{\cal B};\bar F,\lambda \rangle=|{\cal B};\bar F,\lambda,+ \rangle
-|{\cal B};\bar F,\lambda,- \rangle.
\end{equation}
First we construct the boundary state for a D9-brane ($p=9$)
in the absence of the boundary interaction (\ref{coszs}).
We obtain the overlap condition for $\psi_W$ and $\bar\psi_W$
in the closed string channel from (\ref{fbd}),
\begin{equation}
[(1+\hat F)\psi_W+i\epsilon(1-\hat F)\bar\psi_W]\big|_{\tau=0}=0,
\end{equation}
and in terms of the oscillators, this becomes
\begin{equation}\label{bcchi}
(\chi_r + i\epsilon\hat M \bar \chi_{-r}) |{\cal B};\bar F,\epsilon \rangle
= 0,
 \qquad r \in \mathbb{Z}+1/2,
\end{equation}
where $\chi_r$ and $\bar\chi_r$ denote the oscillators of $\psi_W$
and $\bar\psi_W$ respectively.
Solving (\ref{bcchi}) and combining the result in the bosonic case, we have
\begin{equation}
|{\cal B}; \bar F,\epsilon\rangle
=N_p \sqrt{{\bar {\cal Y}}_{p}} \exp\left[ -\sum_{n=1}^\infty
                         \frac1n \sigma_{-n}^T \hat M \bar\sigma_{-n}
-i\epsilon\sum_{r=1/2}^\infty \chi_{-r}^T \hat M \bar\chi_{-r}
                  \right]| {\cal B};\bar F\rangle_0
   \otimes |\mbox{ghost},\epsilon\rangle,
\end{equation}
where $| {\cal B};\bar F\rangle_0$ is the zero mode of the boundary
state.
Since the difference between the bosonic and the superstring case
is only the compactified radius of $W^0$,
we obtain the zero mode of the unperturbed boundary state
by replacing the winding number in the bosonic case
as $m^{(0)}\to \sqrt2m^{(0)}$,
\begin{equation}
| {\cal B};\bar F\rangle_0 = \sum_{m\in \mathbb{Z}}
                  \exp\left[ -i\sqrt2m (w_L^0 - w_R^0)
                     + i\sqrt2m(\hat F_{10} w^1 + \hat F_{20} w^2) \right]
                 |0 \rangle.
\end{equation}
Thus we obtain the boundary state with $\lambda=0$,
\begin{align}
|{\cal B}; \bar F,\epsilon\rangle
= &N_p\sqrt{{\bar {\cal Y}}_{p}} \sum_{j,m} |j;-m,m,\epsilon
\rangle\rangle_{\bar\tau,\bar\delta}^{(0)}\nonumber\\
&\hspace{-12mm}
\otimes \exp\left[ - \sum_{n=1}^\infty\frac1n (\sigma_{-n}^1 \bar\tau_{-n}^1
+ \sigma_{-n}^2 \bar\tau_{-n}^2)
-i\epsilon\sum_{r=1/2}^\infty(\chi_{-r}^1\bar\delta_{-r}^1
+\chi_{-r}^2\bar\delta_{-r}^2)
+ \sqrt2im(\hat F_{10} w^1 + \hat F_{20} w^2) \right]
|0 \rangle \nonumber \\
&\hspace{-12mm}
\otimes_{i=3}^9 |N,\epsilon \rangle_{X^i,\psi^i} \otimes |\mbox{ghost}\rangle,
\end{align}
where $\bar\delta=\hat M\bar\chi$ and
$|j;-m,m,\epsilon\rangle\rangle_{\bar\tau,\bar\delta}^{(0)}$ is the state
obtained
by replacing the $\bar\sigma$ and $\bar\chi$ oscillators by the corresponding
$\bar\tau$ and $\bar\delta$ oscillators on the right-part of the state
appearing in the expansion of $ |j;-m,m,\epsilon\rangle\rangle^{(0)}$.

Let us turn on the boundary interaction (\ref{coszs}).
On the boundary, we see $Z^0=2W_L^0$
and $\Psi_Z^0=\psi_W^0$, and hence the boundary term is expressed
in terms of the left-moving SU(2) current,
\begin{equation}
-i\sqrt2\Psi_Z^0\sin\left(\frac{Z^0(t)}{\sqrt2}\right)\otimes\sigma_{1}
=-i\sqrt2\psi_W^0\sin(\sqrt2W_L^0(t))
\otimes\sigma_{1}= J_{W_L^0}^1(t).
\end{equation}
Then the boundary state with the boundary deformation becomes
\begin{align}
|{\cal B};\bar F,\lambda,\epsilon \rangle
=&\exp( -2\pi i \lambda Q_{W_L^0}^1 ) |{\cal B};\bar F,\epsilon
\rangle \nonumber\\
=& N_p\sqrt{{\bar {\cal Y}}_{p}} \sum_{j=0,1,\ldots}\sum_{m'=-j}^j\sum_{m=-j}^j
 D_{m',-m}^j |j;m',m,\epsilon\rangle\rangle_{\bar\tau,\bar\delta}^{(0)}
\nonumber \\
&\hspace{-15mm}
\otimes \exp\left[ -\sum_{n=1}^\infty\frac1n (\sigma_{-n}^1 \bar\tau_{-n}^1
+ \sigma_{-n}^2 \bar\tau_{-n}^2)
-i\epsilon\sum_{r=1/2}^\infty(\chi_{-r}^1\bar\delta_{-r}^1
+\chi_{-r}^2\bar\delta_{-r}^2)
+ i\sqrt2m(\hat F_{10} w^1 + \hat F_{20} w^2) \right]
|0 \rangle \nonumber \\
&\hspace{-15mm}
\otimes_{i=3}^9 |N,\epsilon \rangle_{X^i,\psi^i} \otimes
|\mbox{ghost},\epsilon\rangle.
\end{align}
Taking the decompactification limit, only the states with $m=m'$
survive and we obtain
\begin{align} \label{sbs}
|{\cal B};\bar F,\lambda,\epsilon \rangle
=&N_p \sqrt{{\bar {\cal Y}}_{p}} \sum_{j=0,1,\ldots}\sum_{m=-j}^j
 D_{m,-m}^j |j;m,m,\epsilon\rangle\rangle_{\bar\tau,\bar\delta}^{(0)}
\nonumber \\
  &\hspace{-15mm}
\otimes \exp\left[ -\sum_{n=1}^\infty\frac1n (\sigma_{-n}^1 \bar\tau_{-n}^1
+ \sigma_{-n}^2 \bar\tau_{-n}^2)
-i\epsilon\sum_{r=1/2}^\infty(\chi_{-r}^1\bar\delta_{-r}^1
+\chi_{-r}^2\bar\delta_{-r}^2)
+ i\sqrt2m(\hat F_{10} w^1 + \hat F_{20} w^2) \right]
       |0 \rangle \nonumber \\
&\hspace{-15mm}
\otimes_{i=3}^9 |N,\epsilon \rangle_{X^i,\psi^i} \otimes
|\mbox{ghost},\epsilon\rangle.
\end{align}

\section{Physical Quantities and Interpretation}
\label{section4}

In section~\ref{section2}, we obtained new
tachyon vertices (\ref{lv}) which are exactly marginal deformations in
the BCFT's for bosonic string and superstrings.
The corresponding boundary states
were constructed through the direct BCFT calculation
and the construction via T-duality in section~\ref{section3}, which
result in the same boundary states.
We shall address physics issues in this section.
In subsection \ref{subsection4-1} we calculate the energy-momentum tensor
and the fundamental string current density.
In subsection \ref{subsection4-2} a plausible interpretation
about the obtained codimension-one configurations is discussed.

\subsection{Energy-momentum tensor and string current density}
\label{subsection4-1}
Given the boundary states (\ref{bt}) and (\ref{sbt}), or equivalently,
(\ref{hestate}) and (\ref{sbs}), it is straightforward to calculate
the corresponding energy-momentum tensors and the current densities
of fundamental strings \cite{Sen:2002in,Mukhopadhyay:2002en}.
From (\ref{bt}), (\ref{Amunu}) and (\ref{sbt}),
we obtain the energy-momentum tensor,
\begin{equation}\label{EMT2}
T^{\mu\nu}={\cal T}_p\sqrt{-{\cal Y}_{p}}\Bigg\{\left.\left[-G^{\mu\nu}
+\frac1{G^{11}}(G^{\mu 1}G^{\nu 1}-\theta^{\mu 1}\theta^{\nu 1}) \right]
f(-\kappa x^{1}) -\frac{1+\hat\lambda}{2G^{11}} (G^{\mu 1}G^{\nu 1}
-\theta^{\mu 1}\theta^{\nu 1})\right.\Bigg\},
\end{equation}
which satisfies the conservation law, $\partial_{\mu} T^{\mu\nu} = 0$,
and the current density of fundamental strings,
\begin{equation}\label{FD}
\Pi^{\mu\nu}={\cal T}_p\sqrt{-{\cal Y}_{p}}\Bigg\{\left.\left[\theta^{\mu\nu}
+\frac1{G^{11}}(G^{\mu 1}\theta^{\nu 1}-G^{\nu 1}\theta^{\mu 1}) \right]
f(-\kappa x^{1})
-\frac{1+\hat\lambda}{2G^{11}}
(G^{\mu 1}\theta^{\nu 1}-G^{\nu 1}\theta^{\mu 1})\right.\Bigg\}.
\end{equation}

A remark is in order. The function $f(x)$ in (\ref{fs})-(\ref{fc})
can have singularities in bosonic case. Explicitly, it diverges at
\begin{equation}
x = \begin{cases}
\rm{sgn}(\lambda) \ln |\sinh\pi\lambda|, & \mbox{(for sinh case),} \\
-\ln|2\pi\lambda|, \mbox{ if }\lambda<0, & \mbox{(for exponential case),} \\
\pm\ln|\sin\pi\lambda|, \mbox{ if }-\frac12<\lambda<0, & \mbox{(for cosh case).}
\end{cases}
\end{equation}
Then it is easy to see that at these singularities the tachyon
has a negative value. (In sinh case, (\ref{fs}) is valid only when
$|\sinh\pi\lambda|<1$ as noted before.)
This is consistent with the fact that the effective tachyon potential
is unbounded from below in the region $T(x)<0$ in the bosonic theory.
On the other hand, the positive region $T(x)>0$ corresponds to the
instability to the decay of an unstable D-brane. For the superstring
case there is no singularity in $f(x)$ and hence $T^{\mu\nu}$ and
$\Pi^{\mu\nu}$ are regular everywhere.

Note that $T^{\mu\nu}$ and $\Pi^{\mu\nu}$ in (\ref{EMT2}) and (\ref{FD})
have essentially the same $x$-dependence, since their local parts are
governed by the function $f(x)$. For later use, we give explicit expressions
for the case of an unstable D2-brane: most of the
components of $T^{\mu\nu}$ and $\Pi^{\mu\nu}$ are constants,
\begin{equation}\label{c}
\frac{\Pi^{01}}{E_{1}}=-\frac{\Pi^{12}}{B}=-\frac{T^{01}}{E_{2}B}
=\frac{T^{11}}{E_{2}^{2}-1}=\frac{T^{02}}{E_{1}B}=-\frac{T^{12}}{E_{1}E_{2}}
=\frac{{\cal T}_{2}(1+\hat\lambda)}{2\sqrt{-{\cal Y}_{2}}}>0,
\end{equation}
while the rest three components are given by the sums of an $x$-dependent piece
$(x=x^1)$ and a constant,
\begin{align}
T^{00}
&=-\frac{{\cal T}_{2}\sqrt{-{\cal Y}_{2}}}{E_{2}^{2}-1}f
+\frac{E_{2}^{2}B^{2}-E_{1}^{2}}{E_{2}^{2}-1}\frac{\Pi^{01}}{E_{1}}
,
\label{t00}\\
T^{22}
&= \frac{{\cal T}_{2}\sqrt{-{\cal Y}_{2}}}{E_{2}^{2}-1}f
-\frac{B^{2}-E_{1}^{2}E_{2}^{2}}{E_{2}^{2}-1}\frac{\Pi^{01}}{E_{1}}
,
\label{t22}\\
\frac{\Pi^{02}}{E_{2}}
&=-\frac{{\cal T}_{2}\sqrt{-{\cal Y}_{2}}}{E_{2}^{2}-1}f
+\frac{B^{2}-E_{1}^{2}}{E_{2}^{2}-1}\frac{\Pi^{01}}{E_{1}} .
\label{pi2}
\end{align}
Here it is important to note that the constant piece of the energy
density $T^{00}$ in (\ref{t00})
is positive, while its $x$-dependent piece
is negative everywhere
since both $C^{11}=-1+E_{2}^{2}$ and
$f(x)$ in (\ref{f0})--(\ref{fc}) are always positive for
the physical region of positive $\lambda$.
This is also the same for the fundamental string charge density $\Pi^{02}$ in
(\ref{pi2}) which is parallel to the codimension-one object.
We will discuss more on this in the subsequent subsection.

\subsection{Physical interpretation of codimension-one objects}
\label{subsection4-2}

In this subsection, we discuss physical properties of the codimension-one
objects found in the previous section by investigating the energy-momentum
tensor and the fundamental string current density. Then we also
compare them to the corresponding kink profiles found
in DBI EFT, NCFT, and BSFT.

Let us first remark that, in the T-duality approach of subsection
\ref{subsection3-1}, we started from rolling tachyons (\ref{chy}).
Nevertheless after a series of transformations the resulting profiles
in (\ref{lv}) are static ones. This formal relationship with
the homogeneous rolling tachyons has also been mentioned in DBI EFT, NCFT, and
BSFT~\cite{Kim:2003in,Kim:2003ma,Banerjee:2004cw,Kim:2004xn,Kim:2006mg}.
However, these objects are different from the rolling tachyons
in that they only exist provided that
the dimension $p$ of the unstable D$p$-branes is larger than one ($p\ge2$)
and that the background electromagnetic field is strong enough for
the condition (\ref{condition}) to be satisfied.

Now we study the detailed properties of the states in the presence
of the tachyon given in (i)--(iii) of (\ref{lv}).
Since there are singularities in bosonic theory as discussed above,
we mainly discuss the superstring case.
Then $f(x)$ in (\ref{sfs})--(\ref{sfc}) is an even function of $\lambda$ and we assume
$\lambda$ is positive. Also for the sake of simplicity
we consider the D2-brane case only.

{\bf (i)} \underline{{\bf hyperbolic sine}} : The tachyon profile in
(i) of (\ref{lv}) is hyperbolic sine type connecting two true
disconnected vacua at $T=\pm\infty$. One may call the monotonically
increasing configuration connecting $T(x=-\infty)=-\infty$ and
$T(x=+\infty)=+\infty$ as a kink and the monotonically decreasing
configuration connecting $T(x=-\infty)=\infty$ and $T(x=+\infty)=-\infty$
as an anti-kink.

As noted before, each quantity in (\ref{t00})--(\ref{pi2}) consists
of an $x$-dependent part and a constant part. The latter comes from
the fluid state of fundamental strings on the brane
in the presence of constant electromagnetic field when the tachyon is
condensed~\cite{Gibbons:2000hf,Sen:2000kd,Kwon:2003qn}.
Because of the condition (\ref{condition}) on the electromagnetic field,
the constant pieces of $T^{00}$ and $\Pi^{02}/E_2$ are all positive.

\begin{figure}[t]
\begin{center}
\scalebox{0.79}[0.79]{\includegraphics{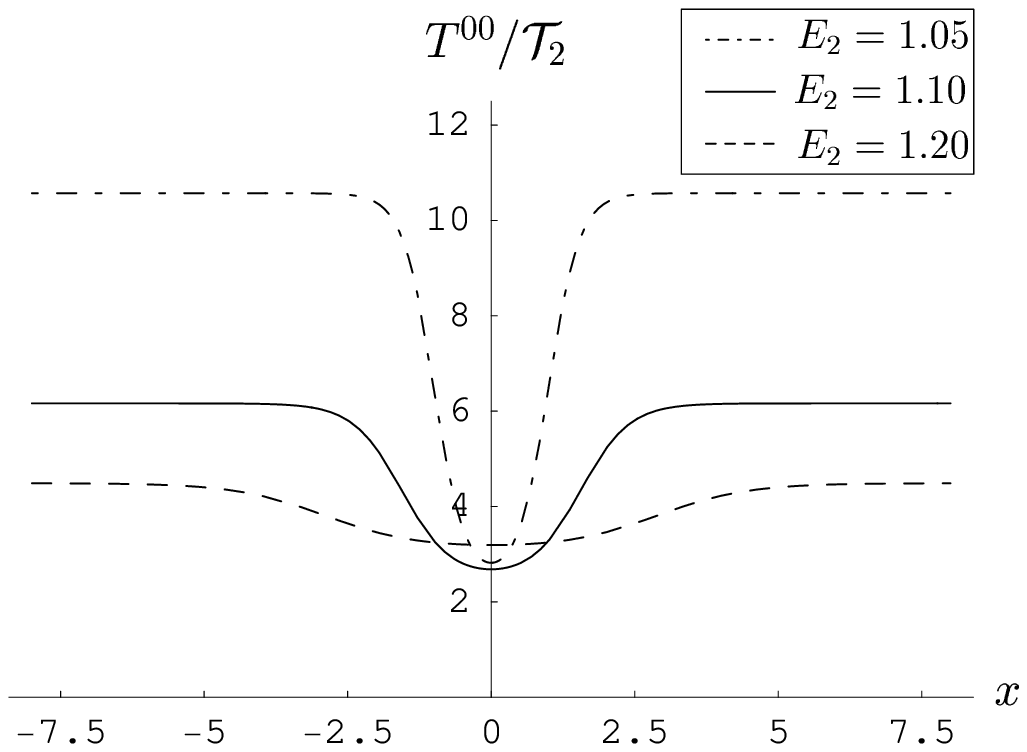}
\includegraphics{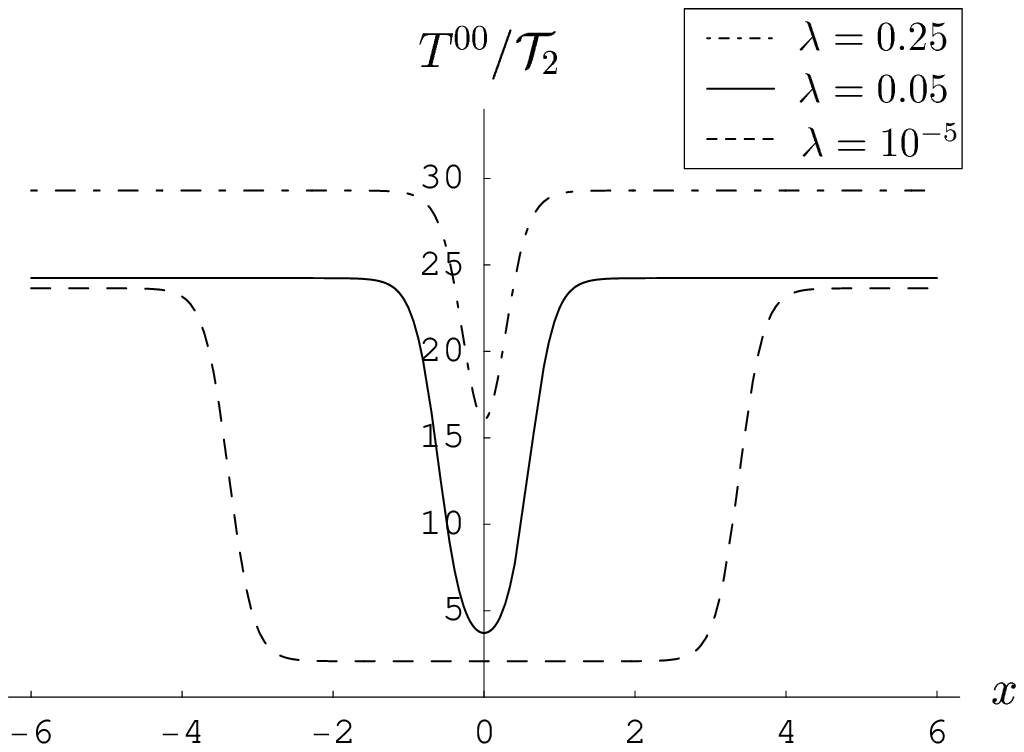}}
\par
\vskip-2.0cm{}
\end{center}
\caption{
{\small The energy density of nonBPS D-brane $T^{00}/{\cal T}_{2}$
with negative tension for various $E_{2}^{2}$ (equivalently
$\Pi^{01}/{\cal T}_{2}$) and $\lambda$.
We choose $E_{1}=0.1$ and $B=0.9$.
The left figure has a fixed $\lambda=0.05$ and three $E_{2}$'s:
$E_{2}=1.05$ ($\Pi^{01}/{\cal T}_{2}$=0.123) for the dot-dashed line,
$E_{2}=1.1$ ($\Pi^{01}/{\cal T}_{2}$=0.133) for the solid line,
and $E_{2}=1.2$ ($\Pi^{01}/{\cal T}_{2}$=0.171)
for the dashed line from the above.
The right figure has a fixed $E_{2}=1.02$
and three $\lambda$'s:
$\lambda=0.15$ for the dot-dashed line, $\lambda=0.05$ for the solid line,
and $\lambda=0.00001$ for the dashed line from the above.
}}
\label{fig2}
\end{figure}

To understand the character of the new objects, we examine the localized
part of the physical quantities (\ref{t00})--(\ref{pi2}).
We consider only the case $\sinh^2\pi\lambda < 1$,
since the form of $f(x)$ can be trusted only for this case \cite{Sen:2004nf}.
Then $f(x)$ is positive definite. It has a maximum at $x=0$ and vanishes
exponentially as $|x| \to\infty$. Since $E_{2}^{2}>1$,
the multiplicative factor in front of the
the localized piece of the energy density in (\ref{t00}) is negative
and it makes a hollow in the energy density.
The depth of the hollow becomes deeper as $E_2^2$ approaches unity as shown in
the left figure of Fig.~\ref{fig2}.
($T^{00}$ can still be shown to be positive definite.)

As $\lambda$ becomes smaller, the hollow becomes almost flat up to
the region $x \sim |\ln \lambda|$ (the right figure of Fig.~\ref{fig2}).
In the vanishing $\lambda$ limit, the energy density at $x=0$ approaches
the value
\begin{equation}
T^{00}(0) \to
 \frac{{\cal T}_2}{\sqrt{-{\cal Y}_2}}\frac{1+B^2}{E_2^2-1},
\end{equation}
which is just the energy density of the unstable D2-brane in the
presence of the electromagnetic field without the deformation,
i.e, for the case $T(x) = 0$. It is quite surprising that
the energy density at $T=0$, which is supposed to be the {\em unstable} point,
is actually smaller than the value at $T=\pm\infty$.
This does not mean, however, that $T=\pm\infty$ has higher vacuum energy
(which is zero) since there is an additional contribution to the energy by
the nonvanishing slope of the tachyon field. Furthermore, the reason that
a hollow is formed is because of the factor $E_2^2-1$ in (\ref{t00})
which flips the sign of the $x$-dependent term to negative when $E_2^2>1$.
This implies that the object has a
negative tension and the accumulated (condensed) constant fundamental
strings in the fluid state are
repelled (de-condensed) along the 1-dimensional brane.
The integration of the localized piece of energy density (\ref{t00})
and the fundamental string charge density (\ref{pi2}) gives
its tension $\tilde{\cal T}_{1}$ and fundamental string
charge per unit length $Q_{{\rm F}1}$ along the $x^{2}$-direction
\begin{equation}
\tilde{\cal T}_{1}=\frac{Q_{{\rm F1}}}{E_{2}}=
-\frac{\sqrt{-{\cal Y}_{2}}}{E_{2}^{2}-1}{\cal T}_{2}
\int_{-\infty}^{\infty} dx f(-\kappa x)
= \frac{\sqrt2{\cal T}_2}{\sqrt{E_{2}^{2}-1}}\ln(\sinh^2\pi\lambda),
\label{te2}
\end{equation}
which is negative as it should be.

According to the profiles of negative energy density,
the configuration near the origin $x=0$ resembles
a hole in condensed matter physics, created in the background of
strong constant electromagnetic field.

{\bf (ii)} \underline{{\bf exponential}} :
For the exponential type of deformation (\ref{lv}), the unstable vacuum at
$T=0$ is connected to the true vacuum at $T=\infty$. In this sense,
one may call the configuration as a half tachyon kink and
an anti-half tachyon
kink respectively for the positive and negative sign in the exponential.
It is convenient to rewrite the corresponding function $f(x)$ in (\ref{sf0}) as
\begin{equation}\label{sfe}
f(\mp\kappa x)=\frac{1}{1+e^{\pm2\kappa(x-x_{0})}},
\qquad \kappa x_{0}=\pm\ln(2\pi\lambda),
\end{equation}
where $x_0$ may be considered as the position of the (anti-)half tachyon kink.

\begin{figure}[ht]
\begin{center}
\scalebox{0.85}[0.85]{\includegraphics{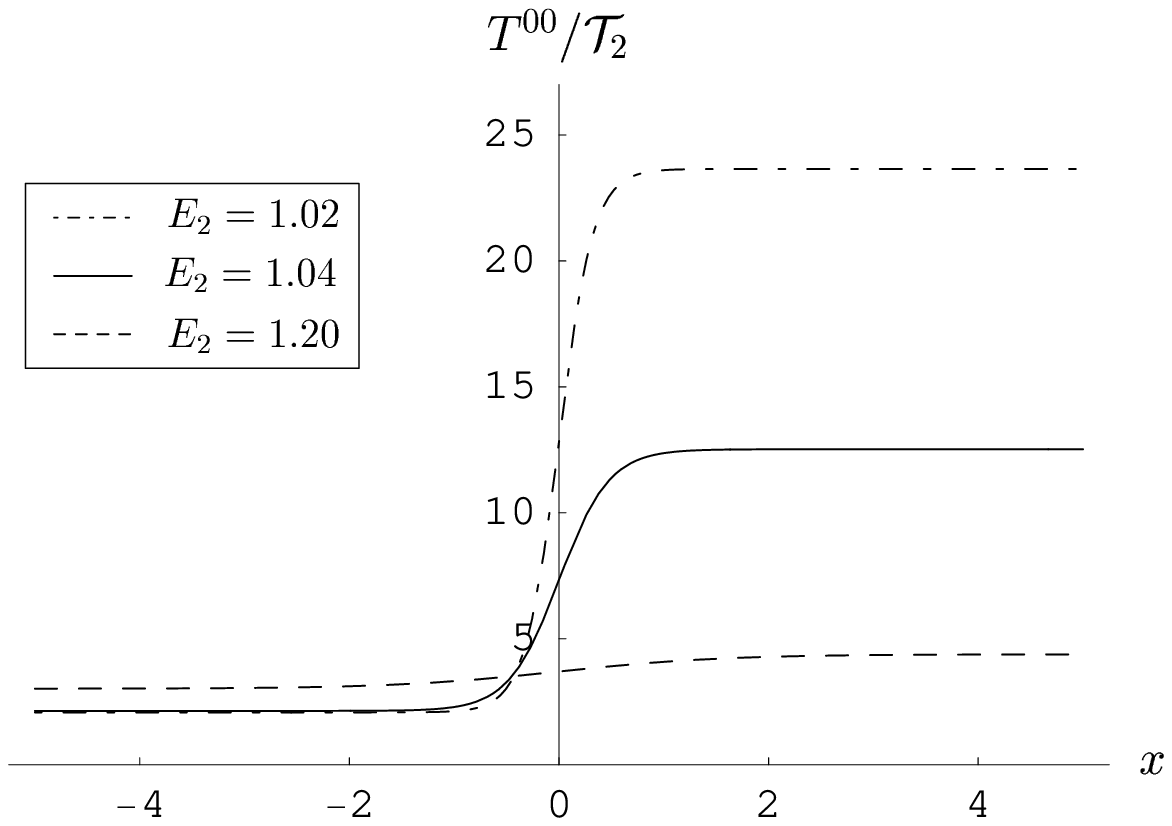}}
\par
\vskip-2.0cm{}
\end{center}
\caption{
{\small The energy density of tensionless half D-brane $T^{00}/{\cal T}_{2}$
for various $E_{2}^{2}$ (equivalently
$\Pi^{01}/{\cal T}_{2}$).
We choose $E_{1}=0.1$ and $B=0.9$, and due to translation, we can choose
$\lambda=1/\sqrt{2\pi}$.
The figure has three $E_{2}$'s:
$E_{2}=1.02$ ($\Pi^{01}/{\cal T}_{2}$=0.115) for the dot-dashed line,
$E_{2}=1.04$ ($\Pi^{01}/{\cal T}_{2}$=0.118) for the solid line,
and $E_{2}=1.2$ ($\Pi^{01}/{\cal T}_{2}$=0.167)
for the dashed line from the above.
}}
\label{fig3}
\end{figure}

From the shape of $f(x)$, we see that the energy density for the half tachyon
kink is monotonically increasing as shown in Fig.~\ref{fig3}.
In this sense the half tachyon kink is a phase boundary stretched along
the $x^{2}$-direction.
In addition the fundamental string charge density $\Pi^{02}$ (\ref{pi2})
is also monotonic.

A naive computation of tension of the half tachyon kink
by integrating the energy
density (\ref{t00}) from $x=-\infty$ to $x=+\infty$ leads to
a divergence due to the contribution from the background energy
proportional to $\int_{x_{0}}^{\infty} dx$.
Therefore, it is reasonable to subtract it
and we obtain
the tension of the half tachyon kink as
\begin{align}
\tilde{\cal T}_{1}&=\int^{\infty}_{-\infty} dx
\left[T^{00}-\frac{E_{2}^{2}B^{2}-E_{1}^{2}}{E_{2}^{2}-1}
\left(\frac{\Pi^{01}}{E_{1}}\right)\right]
-\int_{x_{0}}^{\infty} dx \left(-\frac{{\cal T}_{2}\sqrt{-{\cal Y}_{2}}}{
E_{2}^{2}-1}
\right)
\nonumber\\
&= -\frac{{\cal T}_{2}\sqrt{-{\cal Y}_{2}}}{E_{2}^{2}-1}
\left\{ \int_{x_{0}}^{\infty}dx\left[ \frac{1}{1+e^{2
\kappa (x-x_{0})}}-1 \right] + \int_{-\infty}^{x_{0}}dx
\frac{1}{1+e^{2\kappa (x-x_{0})}}
\right\}\nonumber\\
&=0.
\end{align}
Therefore the half tachyon kink is an object of vanishing energy
and then is identified as a
{\it tensionless half brane} ($\frac{1}{2}$brane) with thickness
$1/2\kappa$. The profile of $\Pi^{02}(x)$ (\ref{pi2}) is almost the same
as the energy density (\ref{t00}).
Suppose the low energy ($T=0$) configuration from $x=-\infty$ and
the high energy ($T=\infty$) configuration from
$x=+\infty$ are glued at $x_{0}$ with a sharp boundary.
Then the formation of a smooth half tachyon kink suggests that
the half phase boundary in the lower energy side gains both energy density
and condensation of the fundamental string charge density, and
the other half phase boundary in the high energy side loses exactly the same
amount of the energy density and condensed fundamental string charge density.
Note that this composite of tensionless $\frac{1}{2}$brane and
fundamental strings has already been obtained
as a half tachyon kink in DBI EFT, NCFT, and BSFT
\cite{Kim:2003in,Kim:2004xn,Kim:2006mg}.

\begin{figure}[t]
\begin{center}
\scalebox{0.81}[0.81]{\includegraphics{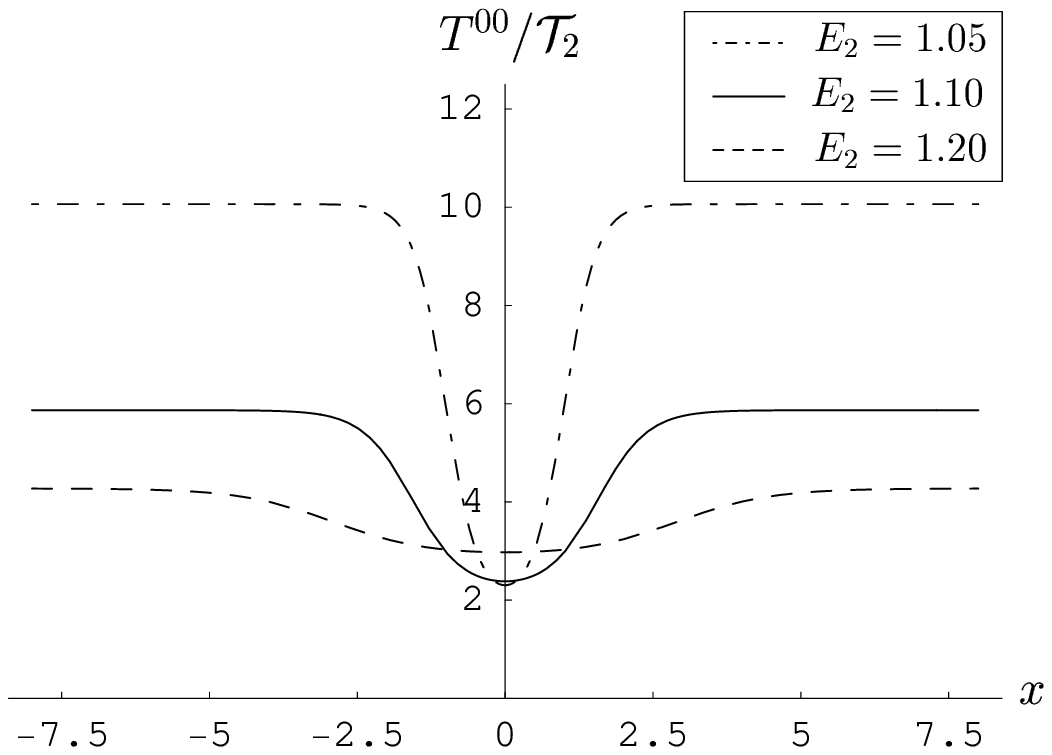}
\includegraphics{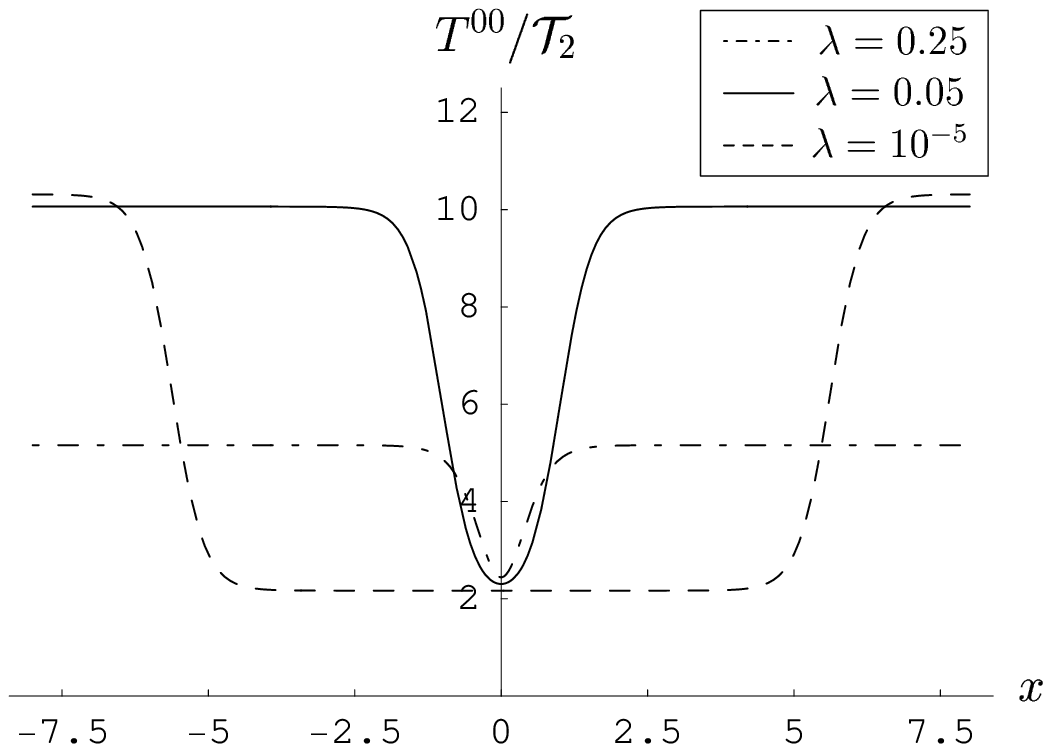}}
\par
\vskip-2.0cm{}
\end{center}
\caption{
{\small The energy density of tachyon bounce $T^{00}/{\cal T}_{2}$
with negative tension for various $E_{2}^{2}$ (equivalently
$\Pi^{01}/{\cal T}_{2}$) and $\lambda$.
We choose $E_{1}=0.1$ and $B=0.9$.
The left figure has a fixed $\lambda=0.05$ and three $E_{2}$'s:
$E_{2}=1.05$ ($-\Pi^{01}/{\cal T}_{2}$=0.117) for the dot-dashed line,
$E_{2}=1.1$ ($-\Pi^{01}/{\cal T}_{2}$=0.127) for the solid line,
and $E_{2}=1.2$ ($-\Pi^{01}/{\cal T}_{2}$=0.163)
for the dashed line from the above.
The right figure has a fixed $E_{2}=1.05$
and three $\lambda$'s:
$\lambda=0.25$ for the dot-dashed line, $\lambda=0.05$ for the solid line,
and $\lambda=0.00001$ for the dashed line from the above.
}}
\label{fig4}
\end{figure}

{\bf (iii)} \underline{{\bf hyperbolic cosine}} :
The hyperbolic cosine type tachyon profile in (iii) of (\ref{lv})
starts from $T=+\infty$, turns at a positive point $\lambda$,
and then goes back to $T=+\infty$,
which may be called a tachyon bounce (or a tachyon breather)
in the classification of solitons.
It could be interpreted as a composite of half tachyon kink and anti-half
tachyon kink. Since the function $f(x)$ (\ref{sfc}) is periodic in $\lambda$,
we may assume $0\le \lambda\le 1/2$.
The shape of the energy density is similar to that of the hyperbolic sine case
as shown in Fig.~\ref{fig4} and we omit the details.

On integrating the localized part of the energy density, we obtain
\begin{equation}
\tilde{\cal T}_{1}=\frac{Q_{{\rm F}1}}{E_{2}}
= \frac{\sqrt2{\cal T}_2}{\sqrt{E_{2}^{2}-1}}\ln(\sin^2\pi\lambda)<0.
\label{desc2}
\end{equation}
This object may be interpreted as a {\it negative-tension
brane} of codimension-one, generated through {\it de-condensing} the
background fundamental strings with a positive constant energy density.
The hyperbolic cosine type tachyon profile also suggests an
interpretation as a composite of half brane and
anti-half brane.
As D${\bar {\rm D}}$ is unstable, this composite of
$\frac{1}{2}$brane and anti-$\frac{1}{2}$brane may also be unstable.
This possible instability is classically
expressed in terms of increase of the minimum
value of the tachyon field in the tachyon bounce configuration.
The parallel component of localized fundamental string charge (\ref{pi2})
is also negative, and it means that the fundamental strings are repelled
at the site of composite as shown in the right graph of Fig.~\ref{fig3}.
Therefore, the obtained object is nothing but a composite
of $\frac{1}{2}$brane and anti-$\frac{1}{2}$brane
accompanying de-condensation of the background fundamental strings.

Let us briefly discuss the bosonic string case.
The cases of exponential
tachyon profile
for positive $\lambda$ (\ref{f0}) and hyperbolic cosine profile (\ref{fc})
for $0<\lambda\le 1/2$ show qualitatively the same regular behavior as
those of superstrings, however the case of hyperbolic sine always involves
singularity.
As mentioned previously,
this can easily be understood
by unbounded nature of the tachyon potential in bosonic string theory
for the region of negative tachyon in (\ref{lv}).

Though we
considered only the case of D1 from an unstable D2, but the extension
to the higher dimensional case of D$(p-1)$ from D$p$
is straightforward.
The aforementioned negative-tension branes and tensionless half brane
are obtained only with a component of overcritical electric field.
As we have discussed, such configurations with overcritical electric field
seem
to be unavoidable in the context of BCFT and may need further study
in the context of string theory beyond BCFT.
Recently the topic of marginal deformations is systematically
dealt in the OSFT~\cite{Schnabl:2007az,Erler:2007rh},
so will be the case of the obtained new tachyon vertices with the development
of constant electromagnetic field.
When the quantum radiations are taken into account including
perturbative closed string modes such as gravitons, dilatons, and
antisymmetric tensor fields, dynamical evolution of the negative-tension
branes and tensionless half brane may become a more intriguing topic.

Finally, let us comment on the results in other approaches
such as DBI EFT and NCFT with 1/cosh tachyon
potential~\cite{Kim:2003in,Kim:2003ma,Kim:2004xn}.
If we compare the physical quantities $T^{\mu\nu}(x)$ and $\Pi^{\mu\nu}(x)$
in the BCFT for superstrings with those in DBI EFT and NCFT with $1/\cosh$
tachyon potential, the BCFT results coincide exactly with the EFT results
for the case of exponential type tachyon
profile ((ii) of (\ref{lv}))~\cite{Kim:2003in,Kim:2003ma,Kim:2004xn}.
For hyperbolic sine ((i) of (\ref{lv})) and
hyperbolic cosine ((iii) of (\ref{lv})) cases, the $x$-dependent part of
physical quantities
match qualitatively but do not exactly coincide with those of DBI
EFT and NCFT.

\section{Conclusions}\label{section5}

In this paper, we considered a flat unstable D$p$-brane ($p\ge 2$)
in the presence of a large constant electromagnetic field in the framework
of BCFT. Specifically, we studied the case that the electromagnetic field
satisfies the following three conditions:
first, a constant electric field is turned on along the $x^1$ direction
($E_{1}\ne 0$); second, the determinant of the matrix $(\eta + F)$ is
negative so that it lies in the physical region ($-\det (\eta + F)>0$);
third, the 11-component of its cofactor is positive
to the large electromagnetic field ($C^{11}>0$).
For the simplest case, $p=2$, these conditions reduce to
$E_{1}\ne 0$, $1-E_{1}^{2}-E_{2}^{2}+B^{2}>0$, and $|E_{2}|>1$, respectively.
In the background of such electromagnetic fields, we identified exactly
marginal deformations depending on the spatial coordinate $x^1$, which
correspond to tachyon profiles of hyperbolic sine, exponential, and
hyperbolic cosine types.
The corresponding boundary states were
constructed by utilizing T-duality approach and also by directly solving
the overlap conditions in BCFT.

For these boundary states, we calculated the energy-momentum tensor and
the fundamental string current density.
Boundary states for the tachyon profiles of hyperbolic
sine, exponential, and hyperbolic cosine correspond to
nonBPS topological kink, half kink, and bounce in the effective field
theories (DBI, NCFT, BSFT), respectively. In superstring theories,
the first and third configurations have
negative tensions and the second configuration gives
tensionless half brane connecting the perturbative string vacuum and
one of the true tachyon vacua.

The result obtained here in the BCFT description completes identifying
all possible codimension-one static solutions on an unstable D$p$-brane
in the presence of a constant electromagnetic field. Without an
electromagnetic field, there exists a unique static solution of which
the tachyon profile is sinusoidal with the period
$2\pi$~\cite{Sen:1998ex,Sen:2004nf}.
When a constant electromagnetic field is turned on,
the spectrum of static solutions becomes rich; there are five types of
solutions for $p\ge 2$, as summarized in the Table~1.
This result coincides with that from DBI type EFT~\cite{Kim:2003in,Kim:2003ma},
NCFT~\cite{Kim:2004xn}, and BSFT~\cite{Kim:2006mg}. The detailed forms of
the energy-momentum tensor and the fundamental string current density
in the BCFT are qualitatively in agreement with those of the EFTs
for cosine (sine), hyperbolic sine, and hyperbolic cosine types of
tachyon profiles. For the linear and the exponential types,
the physical quantities in the BCFT are exactly the same as those
in the EFTs.
\begin{table}[t]
\begin{center}
\renewcommand{\arraystretch}{1.4}
\begin{tabular}{|c | c |c|} \hline
range of electromagnetic field & tachyon profile & interpretation \\ \hline
$C^{11}<0$, $-\det(\eta+F)>0$ & cosine (sine) & array of D$(p-1)({\rm F})
{\bar {\rm D}}(p-1)$(F) \\
$C^{11}<0$, $-\det(\eta+F)\rightarrow 0^{+}$ & linear & single BPS D($p-1$)(F)
\\
$C^{11}>0$, $-\det(\eta+F)>0$ & hyperbolic sine & negative tension brane \\
$C^{11}>0$, $-\det(\eta+F)>0$ & exponential & tensionless half brane \\
$C^{11}>0$, $-\det(\eta+F)>0$ & hyperbolic cosine & negative tension brane\\
\hline
\end{tabular}
\end{center}
\caption{List of all of the static solutions depending on a single spatial
coordinate in the presence of constant electromagnetic field. When the
spatial coordinate is $x^{1}$, the electric field along that direction
should be turned on, $E_{1}\ne 0$.}
\end{table}

\section*{Acknowledgements}
This work was supported by Astrophysical Research
Center for the Structure and Evolution of the Cosmos (ARCSEC)) and
grant No. R01-2006-000-10965-0 from the Basic Research Program
through the Korea Science $\&$ Engineering Foundation (A.I.),
by the Science Research Center Program of the
Korea Science and Engineering Foundation through the Center for
Quantum Spacetime (CQUeST) with grant
number R11-2005-021 and KRF-2006-352-C00010 (O.K.), and
by the Korea Research Foundation Grant funded by the Korean Government
(MOEHRD, Basic Research Promotion Fund, KRF-2006-311-C00022) (C.K.
and Y.K.). A. Ishida, Y. Kim, and O-K. Kwon would like to thank the CQUeST
for the hospitality during various stages of this work.

\appendix
\section{Exponential Type Tachyon Vertex Operator in
BCFT for Superstrings}

In $\sigma$-model approach to string theory, partition function of
the worldsheet action gives the spacetime action and the couplings
are interpreted as spacetime fields~\cite{Fradkin:1985ys}. In
relation to the static tachyon configuration,  exact tachyon
potential and tension of the lower-dimensional D-brane were obtained
from the disk partition function in BSFT~\cite{Gerasimov:2000zp}.
The similar procedure was adapted to the worldsheet action
(\ref{SBCFT}) with exactly marginal tachyon vertex operators. By
identifying the worldsheet partition function with the spacetime
action in bosonic string and superstring theory, the spacetime
energy-momentum tensor without electromagnetic fields was
obtained~\cite{Larsen:2002wc}.

In this appendix we revisit BCFT for superstrings with the
exponential type tachyon profile in (\ref{lv}) by employing the
$\sigma$-model approach. The calculation in bosonic string
theory was given in~\cite{Ishida:2008sp}. According to the procedure
suggested by Ref.~\cite{Larsen:2002wc}, we read the energy-momentum
tensor and fundamental string current density of the unstable system with
exponential type tachyon vertex operator by equating the spacetime
action with the disk partition function of worldsheet theory.

The BCFT for superstrings is distinguished from that for bosonic
string theory by introduction of worldsheet fermions
and the form of
worldsheet action is given in (\ref{Sw}).
Here we use the coordinate $z$ on the unit disk,
\begin{equation}
z=\frac{1+iw}{1-iw}.
\end{equation}
From the action (\ref{Sw}), we again read the worldsheet
energy-momentum tensor
\begin{equation}\label{sTw}
T(z)= -\partial X^\mu
\partial X_\mu (z)-\frac12\psi^{\mu}\partial \psi_{\mu}(z).
\end{equation}

Under the deformed boundary conditions, (\ref{bnd1}) and
(\ref{fbd}), the correlation functions on the unit disk are given by
\begin{align}
\langle X^\mu(z_1)X^\nu(z_2)\rangle_{A}
=&- \eta^{\mu\nu}\ln\mid z_1-z_2\mid
+\eta^{\mu\nu}\ln\mid z_1\bar{z_2}-1\mid
\nonumber \\
&-G^{\mu\nu}\ln\mid z_1\bar{z_2}-1\mid^2
-\theta^{\mu\nu}\ln\left(\frac{z_1\bar{z_2}-1}{\bar{z_1}z_2-1}\right),
\label{XX3}\\
\langle \psi^{\mu}(z)\psi^{\nu}(z^{\prime})\rangle=&
\frac{\eta^{\mu\nu}}{z-z^{\prime}},\qquad\quad
\langle {\bar\psi}^{\mu}({\bar z}){\bar\psi}^{\nu}({\bar
z}^{\prime}) \rangle = \frac{\eta^{\mu\nu}}{{\bar z}-{\bar
z}^{\prime}},
\label{ff2} \\
\langle \psi^{\mu}(z){\bar\psi}^{\nu}({\bar z}^{\prime}) \rangle
=& \left[\frac{1}{\eta-F}(\eta+F)\eta^{-1}\right]^{\mu\nu}
\frac{1}{z-{\bar z}^{\prime}}. \label{ff3}
\end{align}
The two-point function of $\Psi$ at the boundary is the usual one
with the open string metric (\ref{op}),
\begin{equation}\label{ff4}
 \langle \Psi^\mu(t_1) \Psi^\nu(t_2)\rangle=\frac{G^{\mu\nu}}{t_1-t_2},
\end{equation}
where $t_1$ and $t_2$ represent the boundary coordinates on the unit
disk, and the operator product expansion (OPE) for $\psi$ and $\Psi$ becomes
\begin{equation}
\psi^{\mu}(z)\Psi^{\nu}(t)\sim
(G^{\mu\nu}-\theta^{\mu\nu})\frac{1}{z-t}. \label{fb1}
\end{equation}

We turn on a real tachyon field of exponential type,
which is represented by the boundary interaction,
\begin{equation}
S_{T}=
\int_{\partial\Sigma} dt d\theta\, T({\bf X}),\qquad
T({\bf X})=\sqrt2\lambda \exp(i k_\mu {\bf X}^\mu) \otimes \sigma_1,\label{sT}
\end{equation}
where we introduced the superfield ${\bf X}^\mu=X^\mu+i
\sqrt{2}\theta \Psi^\mu$ with boundary Grassmann coordinate $\theta$
and $\sigma_1$ is the Chan-Paton factor in (\ref{TX}). The
corresponding vertex in the zero-picture is obtained by
integrating over $\theta$,
\begin{equation}\label{VT}
 V_T=-2\lambda \,k\cdot \Psi \exp(i k_\mu X^\mu) \otimes \sigma_1.
\end{equation}
The marginality condition for the configurations with single
spatial coordinate dependence (\ref{sk1}) determines the value of $k^\mu$ in
the tachyon vertex operator (\ref{VT}), which is the same as
(\ref{kappa}) for superstrings.

For later convenience, we write the two-point function of the superfield
${\bf X}^1$ on the boundary from (\ref{XX3}) and (\ref{ff3}),
\begin{equation}\label{XXs}
 \langle {\bf X}^1(z_1,\theta_1) {\bf X}^1(z_2,\theta_2)\rangle
=-G^{11}\ln|z_{12}|^2,
\end{equation}
where $z_{12}=z_1-z_2-i\sqrt{z_1 z_2}\theta_1\theta_2$.

From now on we compute the energy-momentum tensor and the fundamental string
current density according to the procedure of
Ref.~\cite{Larsen:2002wc}.
The energy-momentum tensor in BCFT can be read from partition
function of worldsheet theory coupled to background
gravity~\cite{Larsen:2002wc},
\begin{equation}\label{staction}
S=Z_{{\rm disk}} \sim \int[dX][d\psi]e^{-S_{\rm w}-S_T},
\end{equation}
where $Z_{{\rm disk}}$ is the disk partition function, and we
replaced the flat metric $\eta_{\mu\nu}$ to $S_{\rm w}$ with the generic curved
spacetime metric $g_{\mu\nu}$. Then we have the energy-momentum tensor in flat
spacetime:
\begin{equation}
T_{\mu\nu}\equiv -\frac{2}{\sqrt{-g}}\left. \frac{\delta S}{\delta
g^{\mu\nu}} \right|_{g_{\mu\nu}=\eta_{\mu\nu}} =K\sqrt{-\mathcal{Y}_p}
\left[\eta_{\mu\nu}B(x) + A_{(\mu\nu)}(x)\right], \label{EMT}
\end{equation}
where $A_{(\mu\nu)}$ denotes the symmetric part of $A_{\mu\nu}$ and
\begin{align}\label{AB}
 B(x^1)=&-\frac1{\sqrt{-\mathcal{Y}_p}}\int [dX'][d\psi] P \exp(-S_0-S_A-S_T)
=-\langle P \exp(-S_T) \rangle_A,\\
 A^{\mu\nu}(x^1)=&-\frac1{\sqrt{-\mathcal{Y}_p}}\int [dX'][d\psi]
 :(2\partial {X'}^\mu \bar{\partial}
 {X'}^\nu(0)+\psi^\mu \bar{\partial} \psi^\nu(0)
+\bar\psi^\mu \partial \bar\psi^\nu(0)):\nonumber\\
&\quad\times P\exp(-S_0-S_A-S_T)\nonumber\\
=&-\frac1{\sqrt{-\mathcal{Y}_p}}\left\langle :(2\partial {X'}^\mu \bar{\partial}
 {X'}^\nu(0)+\psi^\mu \bar{\partial} \psi^\nu(0)
+\bar\psi^\mu \partial \bar\psi^\nu(0)):
P\exp(-S_T)\right\rangle_A \label{As0}.
\end{align}
Note that we split $X^{\mu}$ into the center of mass coordinate $x^{\mu}$
and fluctuation $X^{'\mu}$, i.e., $X^{\mu}=x^{\mu}+X^{'\mu}$, and
$\langle \cdots \rangle_A$ denotes the vacuum expectation value in
the presence of the U(1) gauge field on the unit disk with normalization
\begin{equation} \label{vev}
\langle 1 \rangle_{A} = \sqrt{-\mathcal{Y}_p}\,.
\end{equation}
In the calculation of $A^{\mu\nu}$ in (\ref{AB}), we follow the
reference \cite{Sen:2002in} and suppose that only the even part
in $\sigma_1$ contributes to the result,
\begin{equation}
\left\langle \cdots P \exp(-S_T)\right\rangle_A =\left\langle
\cdots P \exp(-S_T)\right\rangle_A \Big|_{\sigma_1-{\rm even}}.
\end{equation}

First we calculate $B(x^1)$. Using the two-point function
(\ref{XXs}), we have
\begin{align}
B(x^1)=&-\sum_{n=0}^\infty(2\sqrt2\pi\lambda e^{\kappa
x^1})^{2n} \int \prod_{i=1}^{2n}\frac{dt_i}{2\pi}d\theta_i
\Theta(t_1-t_2)
\cdots \Theta(t_{2n-1}-t_{2n}) \nonumber\\
&\hspace{42mm}\times\prod_{i<j}
|e^{it_i}-e^{it_j}-ie^{\frac{i}{2}(t_i+t_j)\theta_i\theta_j}|
\nonumber\\
=&-\sum_{n=0}^\infty(-4\pi^2\lambda^2 e^{2\kappa x^1})^{n}
\nonumber\\
=&-f(\kappa x^1).\label{Bs}
\end{align}
Then we compute the function $A^{\mu\nu}$. Since the second and third
terms in (\ref{As0}) vanish, (\ref{As0}) becomes
\begin{equation}\label{Amn}
A^{\mu\nu}(x^1) =\frac{-2}{\sqrt{-\mathcal{Y}_p}}\left\langle :\partial X^\mu \bar{\partial}
 X^\nu(0):P\exp(-S_T)\right\rangle.
\end{equation}
For the calculation of $A^{\mu\nu}(x)$ in (\ref{Amn}), we
introduce a different normal ordering for convenience,
\begin{equation}\label{normord2}
\no\partial X^\mu (z) \bar \partial X^\nu (z')\no =
\partial X^\mu (z) \bar \partial X^\nu (z')
 - \partial\bar\partial^{'}
\langle X^\mu (z) X^\nu (z')\rangle_A.
\end{equation}
The relationship between the ordinary normal ordering
and the new ordering (\ref{normord2}) is given by
\begin{equation}\label{normord3}
:\partial X^\mu (0) \bar \partial X^\nu (0): = \no \partial X^\mu
(0) \bar\partial X^\nu (0)\no + G^{\mu\nu} + \theta^{\mu\nu} -
\frac12\eta^{\mu\nu},
\end{equation}
and this leads to
\begin{align}
A^{\mu\nu}(x^1)=&\frac{-2}{\sqrt{-\mathcal{Y}_p}}\left\langle\left( \no\partial X^\mu
\bar{\partial}
 X^\nu(0)\no-\left(\frac{1}{2}\eta^{\mu\nu}-G^{\mu\nu}-\theta^{\mu\nu}
\right)\right)
P\exp(-S_T)\right\rangle\nonumber\\
=&-2\left[\frac1{\sqrt{-\mathcal{Y}_p}}\left\langle\no
\partial X^\mu \bar{\partial}
 X^\nu(0)\no P\exp(-S_T)\right\rangle
-\left(\frac{1}{2}\eta^{\mu\nu}-G^{\mu\nu}-\theta^{\mu\nu}
\right)f(\kappa x^1)\right].
\end{align}
The first term is calculated as follows,
\begin{align}
\frac1{\sqrt{-\mathcal{Y}_p}}&
\left\langle\no\partial X^\mu \bar{\partial}
 X^\nu(0)\no P\exp(-S_T)\right\rangle\nonumber\\
=&\sum_{n=1}^\infty(2\sqrt2\pi\lambda e^{\kappa x^1})^{2n} \int
\prod_{i=1}^{2n}\frac{dt_i}{2\pi}d\theta_i \Theta(t_1-t_2)
\cdots \Theta(t_{2n-1}-t_{2n}) \nonumber\\
&\quad\times \kappa^2 (G^{\mu 1}+\theta^{\mu 1}) (G^{\nu
1}-\theta^{\nu 1}) \left(\sum_{k,l} e^{-t_k-t_l}\right) \prod_{i<j}
|e^{it_i}-e^{it_j}-ie^{\frac{i}{2}(t_i+t_j)\theta_i\theta_j}|
\nonumber\\
=&2\kappa^2 (G^{\mu 1}+\theta^{\mu 1}) (G^{\nu
1}-\theta^{\nu 1}) \sum_{n=1}^\infty
(-4\pi^2\lambda^2 e^{2\kappa x^1})^n\nonumber\\
=&-\frac1{G^{11}}(G^{\mu 1}+\theta^{\mu 1}) (G^{\nu
1}-\theta^{\nu 1}) [f(\kappa x^1)-1].
\end{align}
Finally we obtain
\begin{equation}\label{As}
 A^{\mu\nu}=2\left\{\frac1{G^{11}}(G^{\mu 1}+\theta^{\mu 1})
(G^{\nu 1}-\theta^{\nu 1})[f(\kappa x^1)-1]+\left(\frac{1}{2}\eta^{\mu\nu}
-G^{\mu\nu}-\theta^{\mu\nu}\right)f(\kappa x^1)\right\}.
\end{equation}
As in the bosonic case~\cite{Ishida:2008sp}, we determine the normalization
constant in (\ref{EMT}) as $K={\cal T}_p/2$. Substituting (\ref{Bs})
and (\ref{As})
into (\ref{EMT}), we get the energy-momentum tensor (\ref{EMT2}).
Similarly we obtain the fundamental string current density
(\ref{FD}) which is proportional to the antisymmetric part of
$A^{\mu\nu}$ (\ref{As}).

For the sine and cosine type profiles, the above path integral
method in BCFT is not well-defined due to singularities in OPE
between two vertices. To obtain meaningful results for these cases,
we have to adopt appropriate regularization schemes which await
further development. This situation may have some similarity to the
recently-obtained exactly marginal solutions in
OSFT~\cite{Schnabl:2007az}. The exponential type marginal solution
is well-defined since every OPE between two vertex operators is
regular, while
the sine and cosine type marginal solutions encounter singular behaviors
of OPE between vertex operators.

\end{document}